\documentclass[pdflatex,sn-mathphys-num]{sn-jnl}% Basic Springer Nature Reference Style/Chemistry Reference Style

\usepackage{graphicx}%
\usepackage{multirow}%
\usepackage{amsmath,amssymb,amsfonts}%
\usepackage{amsthm}%
\usepackage{mathrsfs}%
\usepackage[title]{appendix}%
\usepackage{xcolor}%
\usepackage{textcomp}%
\usepackage{manyfoot}%
\usepackage{booktabs}%
\usepackage{algorithm}%
\usepackage{algorithmicx}%
\usepackage{algpseudocode}%
\usepackage{listings}%

\usepackage{mathtools}%
\usepackage{setspace}
\usepackage{lmodern}
\usepackage{anyfontsize}
\usepackage{geometry}
\usepackage{dsfont}

\let\orcidlogo\undefined
\usepackage{orcidlink}

\newgeometry{margin=3.5cm}

\begin{document}

\title{\centering Structural identifiability of partially-observed \\ stochastic processes: from single-particle \\ trajectories to total particle density data}

\author*[1]{\fnm{Arianna} \sur{Ceccarelli}\orcidlink{0000-0002-9598-8845}}\email{ceccarelli@maths.ox.ac.uk}

\author[2]{\fnm{Alexander P.} \sur{Browning}\orcidlink{0000-0002-8753-1538}}

\author[1]{\fnm{Ruth E.} \sur{Baker}\orcidlink{0000-0002-6304-9333}}

\affil[1]{\centering
Mathematical Institute, University of Oxford, UK}%\\Woodstock Road, Oxford, OX2 6GG
\affil[2]{\centering School of Mathematics and Statistics, University of Melbourne, Australia}

\abstract{The increasing availability of experimental data has intensified interest in calibrating stochastic models, raising fundamental questions about parameter identifiability. Structural identifiability determines whether parameters can be uniquely recovered from idealised data, a prerequisite to allow for parameter estimation in real-world scenarios. However, there is a lack of methodologies generally applicable to analyse the structural identifiability of such stochastic processes. Therefore, we develop a methodology for a class of spatio-temporal stochastic processes. We investigate how structural identifiability depends on the type of available data, distinguishing between single-particle trajectories and total particle density measurements. For trajectory data, we use a particle-based description that explicitly represents single-particle dynamics. For population-level data, we derive a partial differential equation representation, that describes the evolution of total particle density, and apply a differential algebra approach, common to ordinary differential equation analysis. We further introduce Taylor-series-expansion and Fourier-transform approaches to study the initial condition. We apply our methodology to an example model with a specified class of initial conditions. We show that the model is globally structurally identifiable given infinitely many, infinitely long single-particle trajectories, but only locally structurally identifiable from total particle density data, demonstrating that parameter identifiability depends on the type of data available.}


\maketitle

%\tableofcontents

\section{Introduction}

The current growth in the quality and quantity of experimental data is leading to an increased interest in calibrating mathematical models to gain quantitative insight. A central question is whether model parameters can be inferred from the available data, referred to as model identifiability. Identifiability analysis plays a crucial role in determining whether meaningful parameter inference is possible, as well as in guiding model formulation, experimental design, and data collection strategies for complex systems.

We focus on structural identifiability, which characterises whether model parameters can be identified in an idealised noiseless setting. A model is \textit{locally structurally identifiable} if its parameters can be determined up to a discrete set of possible values, and \textit{globally structurally identifiable} if parameters can be uniquely determined~\cite{preston2025think,wieland2021structural,walter2014identifiability,cobelli1980parameter,bellman1970structural}. If a model is not locally structurally identifiable, it is \textit{structurally non-identifiable}. For structurally non-identifiable or only locally identifiable models, parameter calibration is ill-posed since even idealised observations of the model output cannot distinguish between these parameters. For structurally identifiable models, the next step is often to analyse their practical identifiability, which concerns finite, noisy data~\citep{preston2025think,wieland2021structural,walter2014identifiability}.

\begin{figure*}[!ht]
    \centering
    \includegraphics[width=\textwidth]{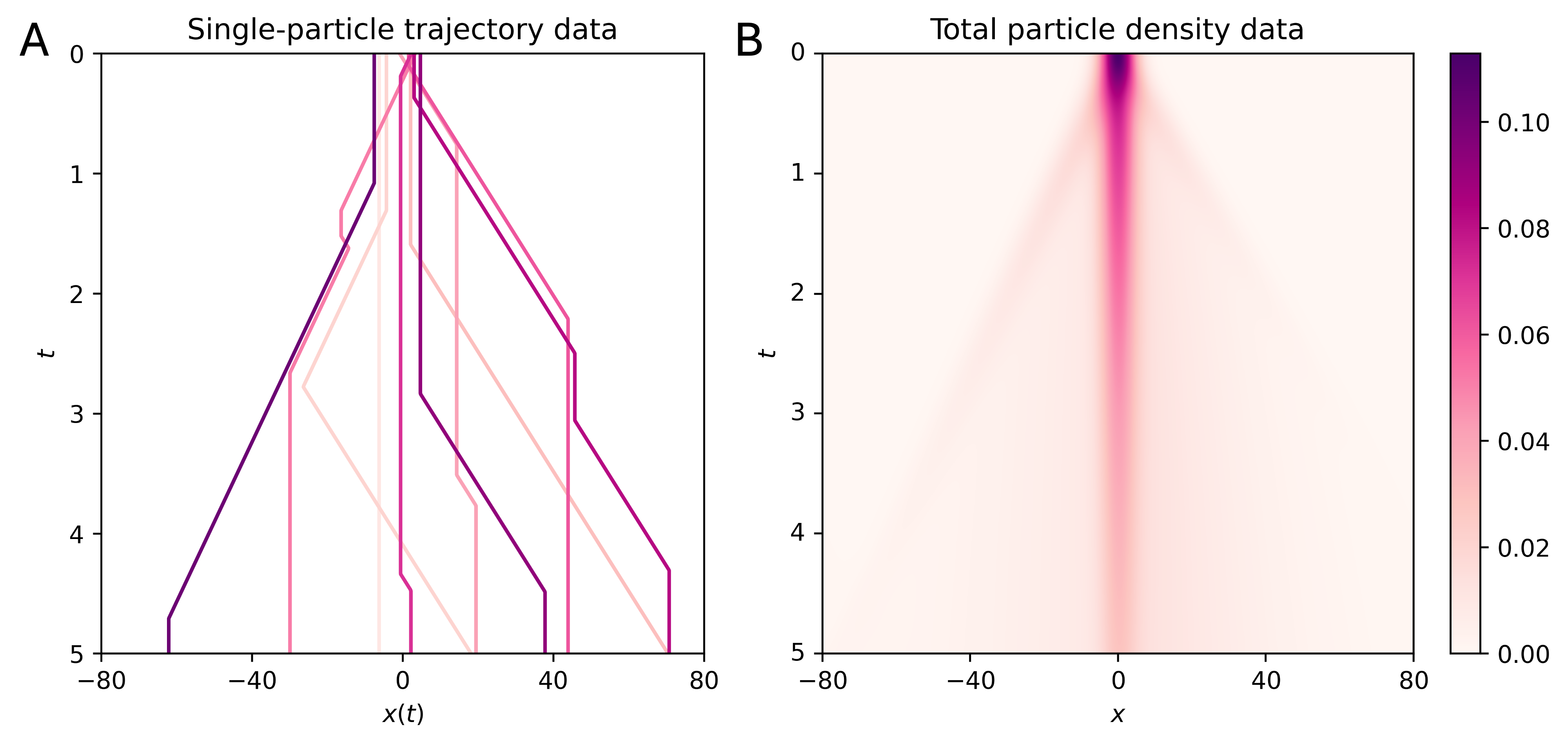}
    \caption{\textbf{Comparison of the information obtained from single-particle trajectory data versus total particle density data.} The model used to generate the data is described in Section~\ref{Sec:models}. \textbf{A} shows examples of single-particle trajectories $x(t)$. \textbf{B} shows an example of information obtained from total particle density data, $N(x,t)$, which gives no access to the evolution of each individual particle.}
    \label{Fig:data}
\end{figure*}

Spatio-temporal stochastic processes can be observed at either the individual or population level. Individual-level data consist of single-particle trajectories, $x(t)$, which capture the location $x$ of individuals, referred to as particles, over time $t$ (Figure~\ref{Fig:data}\textbf{A}; also see~\citep{ceccarelli2026,ceccarelli2025}). Population-level data instead describe the distribution of particles across locations $x$ over time $t$, where the particles evolve according to the same underlying stochastic motion process (Figure~\ref{Fig:data}\textbf{B}).

Structural identifiability analysis for partially-observed stochastic processes is an active area of current research. Specifically, recent work on particle-based models focuses on hidden Markov models with two states~\citep{siekmann2026modelling} and lattice-based random-walk models~\citep{simpson2026trajectories}. However, establishing structural identifiability for such stochastic processes remains challenging due to the limited availability of broadly applicable computational methods.

Moreover, the type of data available, single-particle trajectory data or total particle density data, may lead to specific structural identifiability properties of the model. In particular, we note that, from an infinite number of single-particle trajectories, we can extract total particle density data by obtaining and summing the distributions of particle locations over time for each different state. However, the converse is not true since total particle density data only describe the collective evolution and do not allow to distinguish between individual behaviours and internal states. Hence, the method applied to analyse the model identifiability should also depend on the type of data considered.

We focus on partially-observed stochastic processes in which particle motion depends on an internal Markov state that is not directly measured. We note that, for partially-observed stochastic processes, different hidden-state dynamics may produce the same observed particle motion or density evolution, making structural identifiability non-trivial. For the methods proposed, we consider processes with both a stochastic particle-based model representation, whose simulations directly produce single-particle trajectory data, and a population-level representation based on a differential equation model, whose solution represents the particle densities in each state.

Several methods have been proposed to investigate the structural identifiability of ordinary differential equation (ODE) models~\citep{miao2011identifiability}. These are based on differential algebra and geometry~\citep{villaverde2016structural,raue2014comparison,meshkat2009algorithm,ljung1994global}, profile likelihoods~\citep{raue2009structural}, series expansions~\citep{saccomani2003parameter,pohjanpalo1978system}, or similarity transformations~\citep{vajda1989similarity}; and several software packages have been developed, such as StructuralIdentifiability.jl~\citep{dong2023structidjl}, SIAN~\citep{hong2019SIAN}, GenSSI~\citep{ligon2018genssi,chics2011genssi}, DAISY~\citep{bellu2007daisy}, STRIKE-GOLDD~\citep{diaz2023strike}, and StrikePy~\citep{rey2022strikepy}. 

Structural identifiability methods have also been developed for some stochastic differential equation (SDE) models~\citep{browning2025exact,browning2020identifiability}. For partial differential equation (PDE) models, identifiability has been studied locally using sensitivity-based or information-based approaches~\citep{ciocanel2024parameter,eisenberg2014determining} and globally through extensions of the differential algebra approach for ODEs~\citep{byrne2025algebraic,salmaniw2025structural,browning2024structural,renardy2022structural}.

The differential algebra approach allows to analyse the structural identifiability of ODE and PDE models through the elimination of unobserved state variables from the governing equations, leaving a set of \textit{input-output equations} that depend only on measurable quantities and the model parameters~\citep{renardy2022structural,saccomani2003parameter,audoly2002global,ljung1994global,bellman1970structural}. These equations directly relate the observable outputs to the model parameters, providing a mathematical description of the information contained in the data. Structural identifiability is then assessed by examining the coefficients of the input-output equations, expressed as combinations of the model parameters. If two distinct parameter sets produce the same input-output equations, they give rise to identical observable behaviour and, therefore, cannot be distinguished from any amount of data~\citep{saccomani2003parameter,audoly2002global,ljung1994global,bellman1970structural}. Conversely, if the input-output equations uniquely determine the parameter values, then the model is globally structurally identifiable.

In this work, we propose a methodology to analyse the structural identifiability of a class of partially-observed stochastic processes. We demonstrate the methodology using the stochastic process introduced in~\citep{ceccarelli2026,ceccarelli2025}, for which we derive a PDE representation that enables the application of the differential algebra approach. We first analyse structural identifiability given infinitely many, infinitely long single-particle trajectories, then with total particle density data through the PDE model representation.

To conclude the structural identifiability analysis of the PDE model, we also study the impact of the initial condition. The initial condition influences the model solution and, therefore, can affect structural identifiability~\citep{chis2011structural,ljung1994global,diop1991nonlinear,tunali1987new}. Previous work has obtained an initial-condition input-output equation by evaluating the PDE system at the initial time and applying elimination techniques that characterise the differential algebra approach~\citep{browning2024structural,renardy2022structural}. However, this procedure does not provide a systematic way to derive further initial-condition input-output equations, which may be necessary to fully exploit the information provided by the specified initial condition. Therefore, we develop systematic approaches for deriving the initial-condition input-output equations required for structural identifiability analysis of the PDE models considered here.

Taylor-series-expansion approaches have previously been used to study the structural identifiability of ODE models~\cite{pohjanpalo1978system}. We extend this idea by proposing a novel Taylor-series-expansion approach to extract the initial-condition input-output equations that reveal all identifiable parameter combinations from the form of initial condition considered. We also introduce an alternative approach to obtain the initial-condition input-output equations based on Fourier transforms. Both methods can be used to write the total particle density about the initial time as an ODE, and its derivatives, evaluated at the initial time, give all initial-condition input-output equations.

In Section~\ref{Sec:models} we introduce the stochastic process used to showcase our methods, first presented as a particle-based model, which directly describes single-particle trajectories. Then, we derive a PDE representation of the stochastic model that describes the evolution of the particle densities in each state. In Section~\ref{Sec:SI_SPT}, we analyse the structural identifiability properties of the model with data consisting of an infinite number of single-particle trajectories. In Section~\ref{Sec:SI_density}, we analyse structural identifiability with total particle density data, applying the differential algebra approach to the PDE model. We also incorporate the initial condition in the  structural identifiability analysis using the systematic Taylor-series-expansion and Fourier-transform approaches proposed. Finally, in Section~\ref{Sec:discussion}, we discuss the main contributions of this work, highlighting the differences between the structural identifiability properties of the model with data measuring single-particle trajectory versus total particle density data.

\section{The model and its particle-based versus population-level representations}\label{Sec:models}

\begin{figure*}[!ht]
    \centering
    \includegraphics[width=\textwidth]{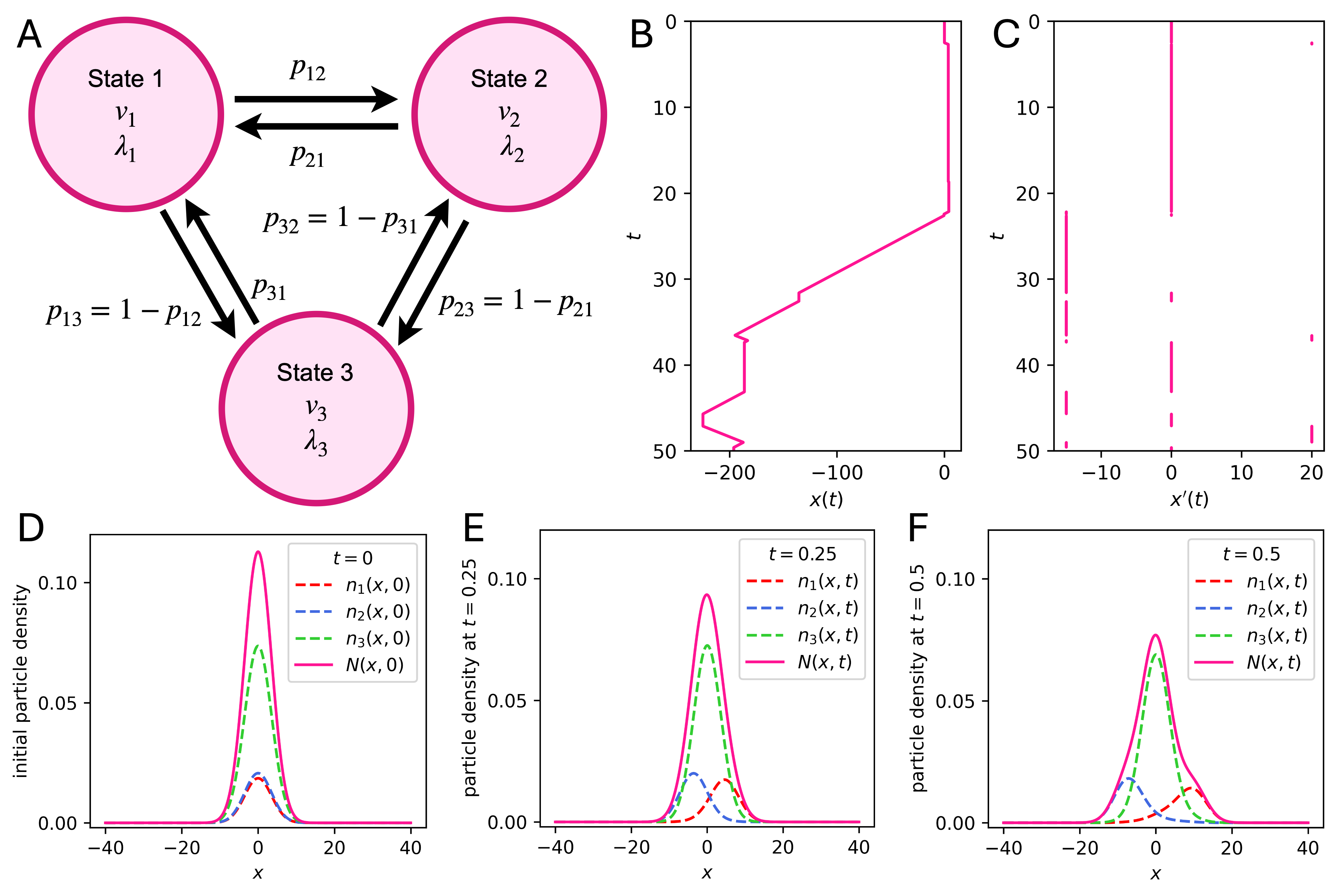}
    \caption{\textbf{The model and the two types of data considered to study its structural identifiability.} \textbf{A} is a graphic visualisation of the three-state particle-based representation. Panels \textbf{B}-\textbf{F} are produced using the parametrised model with parameter set $\boldsymbol{\theta}_A= \{v_1=20, v_2=-15, v_3=0, \lambda_1=1, \lambda_2=1/2, \lambda_3=3/10, p_{12}=1/5, p_{21}=3/10, p_{31}=7/10\}$, denoted model A. \textbf{B} shows a portion of an infinite single-particle trajectory $x(t)$ generated using the parametrised model in \textbf{A}. \textbf{C} shows the piecewise-constant function $x'(t)$, where it exists, obtained from the single-particle trajectory in \textbf{B}. \textbf{D} shows the initial particle densities in each state at time $t=0$, $n_s(x,0)=a_s\exp(-x^2/\varsigma^2)/\sqrt{\pi\varsigma^2}$, for $s=1,2,3$, with $\varsigma=5$ and $[a_1,a_2,a_3]^T=[237/1441, 24/131, 940/1441]^T$ which corresponds to the model stationary distribution (see Supplementary Information Section~S1) (dashed lines), and the total density $N(x,0)$ (continuous line). \textbf{E}-\textbf{F} show the evolution of the particle densities in each state (dashed lines) and the total particle density (continuous line) at time $t=0.25$ and $t=0.5$, respectively, with initial condition in \textbf{D}.}
    \label{Figure_2}
\end{figure*}

In this work, we illustrate our methods by applying them to an example stochastic process for which the structural identifiability properties of have not to-date been studied. We consider a stochastic velocity-jump model in one spatial dimension, in which the particle's internal state evolves as a continuous-time Markov chain within a network of three states~\citep{ceccarelli2026,ceccarelli2025}. We assume that each state is characterised by a constant velocity and fixed rates of switching to every other state; thus, the particle's state evolution fully characterises its motion~\citep{ceccarelli2026,ceccarelli2025} (Figure~\ref{Figure_2}\textbf{A}). Velocity-jump models have been used to describe motion in several contexts, for example, microtubular transport along the axons of neurons~\citep{bressloff2021queuing,xue2017recent,encalada2014biophysical,bressloff2013stochastic,kuznetsov2011analytical,jung2009modeling,friedman2005model,brown2000slow}, cell, animal or bacterial motility and chemotaxis~\citep{taylor2015birds,treloar2011velocity,erban2004individual,othmer1988models}, and swarm robotic motion~\citep{franz2016hard,taylor2015mathematical}. 

\subsection{The stochastic particle-based model representation}\label{Subsec:IBM}

We first consider the stochastic particle-based version of the model, which directly describes the motion of a single particle in one spatial dimension characterised by three internal states, as described in~\citep{ceccarelli2026,ceccarelli2025} (Figure~\ref{Figure_2}\textbf{A}). Each state $s=1,2,3$ is associated with a fixed velocity $v_s\in\mathbb{R}$, not all necessarily distinct, an exponential state-switching process with rate $\lambda_s>0$, and fixed transition probabilities to any other state $u\ne s$, denoted $p_{su}\in[0,1]$, such that
$p_{12}+p_{13}=1$, $p_{21}+p_{23}=1$, and $p_{31}+p_{32}=1$. These probabilities give the switching probability matrix $\boldsymbol{P}=[p_{su}]$, defined with zero diagonal entries. Hence, the model considered has a set of nine parameters to identify, $\boldsymbol{\theta}=\{v_1,v_2,v_3,\lambda_1,\lambda_2,\lambda_3,p_{12},p_{21},p_{31}\}$.

The location $x$ evolves over time $t$, for a particle $m$ according to its internal state, and is represented as a continuous and piecewise-differentiable function $x_m(t)$. The state evolution is a continuous-time Markov chain, and we define its transition matrix $\boldsymbol{Q}$ in terms of the total switching rates $\lambda_s$ and the transition probabilities $p_{su}$ as 
\begin{equation}\label{Eq:Q}
\boldsymbol{Q} = [q_{su}] :=
\begin{bmatrix}
    -\lambda_1 & \lambda_1 p_{12} & \lambda_1 p_{13} 
    \\
    \lambda_2 p_{21} & -\lambda_2 & \lambda_2 p_{23} 
    \\
    \lambda_3 p_{31} & \lambda_3 p_{32} & -\lambda_3 
\end{bmatrix}=
\begin{bmatrix}
    -\lambda_1 & \lambda_1 p_{12} & \lambda_1 (1-p_{12})
    \\
    \lambda_2 p_{21} & -\lambda_2 & \lambda_2 (1-p_{21}) 
    \\
    \lambda_3 p_{31} & \lambda_3 (1-p_{31}) & -\lambda_3
\end{bmatrix},
\end{equation}
such that the chain is irreducible (see~\citep{ceccarelli2026,ceccarelli2025}). Finally, we assume that each particle's initial location is sampled according to a probability distribution $f(x)\ge0$ and its initial state in each location is distributed according to a probability vector $\boldsymbol{a}=[a_1,a_2,a_3]^T$. %In the next section, we describe the PDE system representation of this stochastic process.

\subsection{The population-level system representation}\label{Subsec:PDE_model}

In order to study the structural identifiability with total particle density data, we write the three-state Fokker-Planck PDE model~\citep{gardiner2021markov} description of the evolution of the particle density in each state $s$, denoted $n_s(x,t)$, at location $x\in\mathbb{R}$ and at time $t>0$, given by
\begin{subequations}\label{Eq:RA}
\begin{align}
    &\partial_t n_1+v_1\partial_x n_1= -\lambda_1n_1+\lambda_2 p_{21}n_2+\lambda_3 p_{31}n_3,\label{Eq:a}
    \\
    &\partial_t n_2+v_2\partial_x n_2= \lambda_1 p_{12}n_1-\lambda_2n_2+\lambda_3 (1-p_{31})n_3,\label{Eq:b}
    \\
    &\partial_t n_3+v_3\partial_x n_3= \lambda_1 (1-p_{12})n_1+\lambda_2 (1-p_{21})n_2-\lambda_3 n_3,\label{Eq:c}
\end{align}
\end{subequations}
where $\partial_t$ and $\partial_x$ denote the partial derivatives in time and space, respectively (see Supplementary Information Section~S2 for a full derivation).

As for the particle-based model in Section~\ref{Subsec:IBM}, we assume the initial location of each particle to be sampled according to a probability distribution function $f(x)\ge0$, and the state according to the probability vector $\boldsymbol{a}=[a_1,a_2,a_3]^T$, with $a_s\in[0,1]$ and $a_1+a_2+a_3=1$. Hence, the initial density in each state $s=1,2,3$ is
\begin{equation}\label{Eq:IC}
    n_s(x,0)=a_s f(x),
\end{equation}
where $f(x)=N(x,0)$ is the initial total particle density. For the numerical examples, we choose the initial condition to be a Gaussian around zero, $f(x)=\exp(-x^2/\varsigma^2)/\sqrt{\pi\varsigma^2}$. We consider the spatial domain to be sufficiently large that the particle density becomes negligible far from the region of interest. We therefore idealise the spatial domain as unbounded and impose far-field boundary conditions
\begin{equation}\label{Eq:BC}
    \lim_{|x|\to +\infty}n_s(x,t)=0,
\end{equation}
for $s=1,2,3$. These conditions correspond to assuming that the particle population remains spatially localised, such that the density in each state vanishes sufficiently far from the region occupied by the particles.

For $f$ analytic, the model solution exists and is unique~\citep{godlewski1991hyperbolic}. Indeed, for any parameter choice, the PDE system has all real eigenvalues $v_1,v_2,v_3$, not necessarily distinct, hence, it is \textit{strongly hyperbolic} and strong hyperbolicity guarantees that solutions to the Cauchy problem exist, are unique, and depend continuously on the initial data~\citep{godlewski1991hyperbolic}.

\section{The model parameters are structurally identifiable measuring single-particle trajectories}\label{Sec:SI_SPT}

Now, we propose a novel method to analyse the structural identifiability properties of the particle-based model defined in Section~\ref{Sec:models}, assuming that we measure infinitely many, infinitely long single-particle trajectories denoted $x_1(t), x_2(t), \ldots, x_m(t),\ldots$ (Figure~\ref{Fig:data}\textbf{A}). We initially consider a single infinitely long trajectory denoted $x(t)$. For a velocity-jump process, a trajectory $x(t)$ is continuous but only piecewise differentiable. Therefore, its derivative $x'(t)$ is defined almost everywhere and is piecewise constant, taking values in the set $\{v_1,v_2,v_3\}$ (Figure~\ref{Figure_2}\textbf{C}). Hence, all distinct velocities can be directly identified.

If all velocities are distinct, we can obtain the lengths of all time intervals with constant velocity $v_s$, denoted $\tau_s^1,\tau_s^2,\ldots$, which are associated with state $s$. Then, to obtain $\lambda_s$, we write the cumulative waiting-time distribution of the time spent in each state $s=1,2,3$, as 
$$\mathbb{P}(t_s\le t)=\lim_{K\to\infty}\frac{1}{K}\sum_{k=1}^K \mathds{1} (\tau_s^k\le t),$$
where $t_s$ is the time spent in state $s$ before switching and $\mathds{1}$ is the indicator function~\citep{ross2014introduction}. Then, the survival probabilities can be written as
$$\mathbb{P}(t_s> t)=1-\mathbb{P}(t_s\le t)=\exp(-\lambda_s t),$$
making the switching rates $\lambda_s$ identifiable as 
$$\frac{\text{d}\exp(-\lambda_s t)}{\text{d} t}/\exp(-\lambda_s t)=-\lambda_s.$$

The transition probability from state $s$ to state $u$ can be obtained as
$$p_{su}=\lim_{t\to\infty}\frac{W_{su}(t)}{W_s(t)},$$
where $W_{su}(t)$ denotes the measured number of switches from state $s$ to $u$ in the time interval $[0,t]$, and $W_s(t)$ denotes the total number of switches from state $s$ to any other state in the time interval $[0,t]$ for the trajectory $x(t)$. Finally, we analyse the identifiability of the initial condition. From a single trajectory $x(t)$, we can only identify the particle's initial location $x(0)$ and state $S(0)$ from its initial velocity, 
$$x'(0):=\lim_{t\to 0^+} \frac{x(t)-x(0)}{t}.$$ 

Now, we show that, from an infinite number of trajectories, $x_1(t),x_2(t),\ldots$, we can identify the probability density function of the initial location $f(x)$ and vector $\boldsymbol{a}=[a_1,a_2,a_3]^T$. First, we can obtain the cumulative waiting-time distribution of the initial location as
$$F(x)=\lim_{M\to\infty}\frac{1}{M}\sum_{m=1}^M \mathds{1} (x_m(0)\le x),$$
and the probability distribution function is obtained as $f(x)=F'(x)$. Moreover, we can obtain the particle's initial state probability vector $\boldsymbol{a}=[a_1,a_2,a_3]^T$ as the proportion of particles in each state $s$, with velocity $v_s$, at time $t=0$, writing 
$$a_s = \lim_{M\to\infty} \frac{1}{M} \sum_{m=1}^M \mathds{1} (x_m'(0)=v_s),$$
excluding the particles for which $x_m$ is not differentiable at zero.

In summary, all model parameters, including the initial condition, are structurally identifiable using an infinite number of single-particle trajectories if all states have distinct velocities. In Supplementary Information Section~S3, we analyse the structural identifiability properties of a three-state model in which two states share the same velocity. This case is more complex as the internal state is not distinguishable. Without loss of generality, we assume $v_2=v_3$, and we obtain that the model parameters $v_1,v_2,v_3,\lambda_1$ are identifiable, while the parameters $\lambda_2,\lambda_3,p_{12},p_{21},p_{31}$ are not, but their combinations \begin{equation}\label{Eq:k_1k_2k_3}
    k_1=\lambda_2+\lambda_3, \qquad k_2=\lambda_2 \lambda_3 (p_{21}+p_{31}-p_{21}p_{31}),\qquad k_3=\lambda_2 p_{12} p_{21}+\lambda_3 p_{31}(1 - p_{12}),
\end{equation} 
are identifiable and, analysing the initial condition, $a_1$ and the parameter combination
\begin{equation}\label{Eq:k_4}
    k_4=(1-a_1)\lambda_3 p_{31} + a_2(\lambda_2 p_{21} - \lambda_3 p_{31}),
\end{equation}
are identifiable, but not $a_2$. We conclude that the model with two equal velocities is non-identifiable as the four identifiable parameter combinations $k_1,k_2,k_3,k_4$ cannot be sufficient to identify all six remaining model parameters $\lambda_2,\lambda_3,p_{12},p_{21},p_{31},a_2$.

\section{Structural identifiability measuring the total particle density can be assessed by formulating a PDE model}\label{Sec:SI_density}

In this section, we study the model structural identifiability measuring the total particle density $N(x,t)$ using its PDE formulation obtained in Section~\ref{Subsec:PDE_model}. We note that if $N(x,t)$ is measured, then its time and space derivatives are also observed. First, we apply the differential algebra approach to reduce the PDE system to \textit{input-output equations}, based on differential elimination techniques to remove unobserved state variables from the governing equations~\citep{byrne2025algebraic,salmaniw2025structural,browning2024structural,renardy2022structural}. Then, the input-output equations involve only measured quantities, and their coefficients, written as model parameter combinations, are structurally identifiable when written in a monic form, i.e. with one coefficient set to one~\citep{saccomani2003parameter,audoly2002global,ljung1994global,bellman1970structural}. Finally, the initial condition introduces further unknown functions or parameters and is known to affect the identifiability properties of PDE systems, thus, we incorporate it into the analysis by proposing and applying a novel method to study it.

\subsection{Structural identifiability analysis using the differential algebra approach}\label{Subsec:DA}

We apply the differential algebra approach which, for linear PDE systems, allows to obtain input-output equations by repeatedly differentiating the governing equations and combining them to form a larger PDE system, which can then be solved to eliminate the unobserved variables~\citep{browning2024structural,renardy2022structural}. Now, we assume that only the total particle density $N(x,t)$ is measured, therefore, we aim to eliminate the unobserved variables $n_s(x,t)$, and obtain input-output equations only in terms of $N(x,t)$, its derivatives, and the model parameters. Firstly, we remove the unobserved variable $n_1=N-n_2-n_3$ from Equation~\eqref{Eq:RA} to obtain
\begin{equation}\label{Eq:RA_sum_no_n1}
    \partial_t N+v_1\partial_x N+(v_2-v_1)\partial_x n_2+(v_3-v_1)\partial_x n_3= 0.
\end{equation}
We proceed under the assumption that at least two velocities are distinct, without loss of generality $v_1\ne v_2,v_3$, so that we can fix a state labelling, for example, such that $v_1>v_2$. Under this assumption, we can divide by $v_1-v_2\ne 0$ and use Equation~\eqref{Eq:RA_sum_no_n1} to obtain
\begin{equation}\label{Eq:three-state_n2}
    \partial_x n_2=\frac{\partial_tN+v_1\partial_xN+(v_3-v_1)\partial_x n_3}{v_1-v_2}.
\end{equation}
We then differentiate Equation~\eqref{Eq:b} and Equation~\eqref{Eq:c} with respect to $x$, and substitute Equation~\eqref{Eq:three-state_n2} and its derivatives to obtain equations that involve derivatives of $n_3$ up to order two, and can be used to write a $6\times 6$ linear system in $n_3$ and its derivatives. The system can be reduced to obtain the input-output equation
\begin{equation}\label{Eq:IO}
\partial_{ttt}N+c_1\partial_{xtt}N+c_2 \partial_{xxt}N+c_3 \partial_{xxx}N+c_4\partial_{tt}N+c_5 \partial_{xt}N+c_6 \partial_{xx}N+c_7\partial_t N+c_8 \partial_x N=0,
\end{equation}
with coefficients
\begin{equation}\label{Eq:RA_COEFFS}
\begin{aligned}
    &c_1=v_1 + v_2 + v_3,\qquad c_2=v_1 v_2 + v_2 v_3 + v_3 v_1,\qquad c_3=v_1 v_2 v_3,
    \\
    &c_4=\lambda_1 +\lambda_2 + \lambda_3,\qquad c_5= \lambda_1 (v_2 + v_3) + \lambda_2 (v_1 + v_3)+\lambda_3 (v_1 + v_2),
    \\
    &c_6= \lambda_1 v_2 v_3 + \lambda_2 v_1 v_3 +\lambda_3 v_1 v_2,
    \\
    & c_7=\lambda_2 \lambda_3 (1-(1-p_{21})(1-p_{31})) + \lambda_1 \lambda_3 (1 - (1 - p_{12}) p_{31}) + \lambda_1 \lambda_2 (1 - p_{12} p_{21}), 
    \\
    & c_8=\lambda_2 \lambda_3 (1-(1-p_{21})(1-p_{31})) v_1 + \lambda_1 \lambda_3 (1 - (1 - p_{12}) p_{31}) v_2 + \lambda_1 \lambda_2 (1 - p_{12} p_{21}) v_3.
\end{aligned}
\end{equation}
The calculations to reduce the system to the input-output equation are performed using Mathematica (see the notebook \textit{Section\_4.1\_coefficients\_c1-c8.nb}). Since the input-output equation is monic in $\partial_{ttt}N$, it has a unique normalised representation. Consequently, any two parameter sets that generate the same input-output equation must have identical coefficients, implying that the coefficients $c_1,\ldots,c_8$ are structurally identifiable.

From the first three coefficients, $c_1,c_2,c_3$, we obtain a system of three equations of degree three in three unknowns $v_1,v_2,v_3$. We can write $v_3$ as $v_3=c_1-v_1-v_2$, which gives the identifiability of $v_3$ once $v_1$ and $v_2$ are identified. Substituting the expression for $v_3$ into $c_2,c_3$ we obtain the conic $c_2=v_1 v_2 + v_2 (c_1-v_1-v_2) + (c_1-v_1-v_2)v_1$ and the cubic $c_3=v_1v_2(c_1-v_1-v_2)$, respectively. The possible solutions for $v_1,v_2$ are obtained from the intersection of the conic and the cubic. By Bézout's theorem, the intersection has six zeros $(v_1,v_2,c_1-v_1-v_2)$ counted with multiplicity~\citep{fulton1969algebraic}. By the symmetry of the system, we note that, since $(v_1,v_2,v_3)$ must be a solution of the system given by $c_1,c_2,c_3$, thus, any permutation of those is also a solution. Hence, all six solutions of the system are permutations of $(v_1,v_2,v_3)$. We conclude that all velocities are identifiable up to state relabelling.

Once the state labelling is fixed, the coefficients $c_4,c_5,c_6$ form a linear system in the switching rates $\lambda_1,\lambda_2,\lambda_3$ written as
\begin{equation}\label{Eq:sys_lambda}
    \begin{bmatrix}
    c_4 \\ c_5 \\ c_6
\end{bmatrix}=\begin{bmatrix}
    1 & 1 & 1
    \\
    v_2+v_3 & v_1+v_3 & v_1+v_2
    \\
    v_2 v_3 & v_1 v_3 & v_1 v_2
\end{bmatrix}\begin{bmatrix}
    \lambda_1 \\ \lambda_2 \\ \lambda_3
\end{bmatrix}.
\end{equation}
The matrix on the right hand side has determinant $(v_1-v_2)(v_2-v_3)(v_1-v_3)$. If the velocities are all distinct, this determinant is non-zero, giving a unique solution for the switching rates vector $[\lambda_1,\lambda_2,\lambda_3]^T$, making all rates structurally identifiable. In Supplementary Information Section~S4, we consider the case $v_2=v_3$ and obtain that the velocities $v_1,v_2,v_3$ and the rate $\lambda_1$ remain identifiable from the input-output equation, while the other parameters are not identifiable, but their combinations $k_1,k_2,k_3$, defined in Equation~\eqref{Eq:k_1k_2k_3}, are identifiable.

The three probability parameters $p_{12}, p_{21}, p_{31}$ appear only in the two quadratic equations given by fixing the coefficients $c_7$ and $c_8$. As these are two equations in three unknowns, then the probability parameters are not fully identifiable from the input-output equations by the Implicit Function Theorem~\citep{rudin1976principles}. We note that, so far, we have not assumed any form of the initial condition, and in this case, the model is structurally non-identifiable. Specifically, the velocities $v_s$ and rates $\lambda_s$ are identifiable, but the switching probability parameters are not. In contrast, in Section~\ref{Sec:SI_SPT}, we found that the model with velocities all distinct is structurally identifiable given single-particle trajectory data. Next, we incorporate initial conditions of the form specified in Equation~\eqref{Eq:IC} in the structural identifiability analysis.

\subsubsection*{Incorporating separable initial conditions}

Now, we consider the separable initial conditions of the form specified in Equation~\eqref{Eq:IC}, that assume that the initial states are in the same proportion across all locations. We check whether the coefficients $a_1,a_2,a_3$ are identifiable and if they impact the identifiability of the other model parameters. We note that, since $a_1+a_2+a_3=1$, the function $f(x)$ is identifiable since
\begin{equation}\label{Eq:RA_IC_IO_base0}
    N(x,0)=(a_1+a_2+a_3)f(x)=f(x),
\end{equation} 
is measured; thus its spatial derivatives are also observed. We note that Equation~\ref{Eq:RA_IC_IO_base0} gives the first Cauchy data $N(x,0)$, but does not add any additional information on the model parameters.

Moreover, following the procedure in~\citep{browning2024structural,renardy2022structural}, we substitute the initial condition in Equations~\eqref{Eq:IC} into Equations~\eqref{Eq:RA}, and their sum gives the input-output equation
\begin{equation}\label{Eq:IC_IOeq}
\partial_tN(x,0)+c_9f'(x)= 0,
\end{equation}
where
\begin{equation}\label{Eq:c9_IC}
    c_9=a_1v_1+a_2v_2+a_3v_3=a_1(v_1-v_3)+a_2(v_2-v_3)+v_3.
\end{equation}
First, we note that, since the coefficients $c_1,c_2,c_3$ give structural identifiability of the velocities $v_1,v_2,v_3$, the coefficient $c_9$ fixes a line of possible parameters $(a_1,a_2)$. Therefore, $c_9$ is not sufficient to identify the initial condition parameters $a_1,a_2$.

We note that, the three-state input-output equation in Equation~\eqref{Eq:IO} is third-order in time. Hence, for fixed model parameters, its solution for the total particle density is fixed by three Cauchy data~\citep{strauss2007partial}, for example, $N(x,0)$, $\partial_tN(x,0)$, and $\partial_{tt}N(x,0)$. Indeed, once $N(x,0)$, $\partial_tN(x,0)$, and $\partial_{tt}N(x,0)$ are fixed, further differentiations of Equation~\eqref{Eq:IO} with respect to time then recursively determine all higher-order time derivatives of $N$ at $t=0$.

In Supplementary Information Section~S5.1, we show examples that highlight that the coefficients \texorpdfstring{$c_1,\ldots,c_9$}{c1,...,c9} are not sufficient to establish all structurally identifiable parameter combinations from the separable initial condition, therefore, further initial-condition input-output are required. However, it is unclear how to obtain an expression for the remaining Cauchy datum, $\partial_{tt}N(x,0)$, using the differential algebra approach. We address this issue in Sections~\ref{Subsec:Taylor_IC}-\ref{Subsec:Fourier_IC}, by providing two alternative approaches to find the remaining input-output equation.

\subsection{A Taylor-series-expansion approach for the derivation of the initial-condition input-output equations}\label{Subsec:Taylor_IC}

So far, we have obtained that, if the velocities are all distinct, the velocities and the switching rates are identifiable, but the coefficients $c_1,c_2,\ldots,c_9$ alone do not allow identification of the probability parameters $p_{12},p_{21},p_{31}$ or the initial condition parameters $a_1,a_2$. A classical approach to structural identifiability analysis of ODE models is based on the Taylor-series expansion of the model output, which examines how the coefficients of successive output derivatives depend on the unknown parameters~\citep{pohjanpalo1978system}. In this section, we propose a Taylor-series-expansion approach to derive the input-output equation for $\partial_{tt}N(x,0)$ associated with the specified separable initial condition. More generally, this approach allows systematic derivation of all the initial-condition input-output equations for the model, including the previously obtained quantities $N(x,0)$ and $\partial_t N(x,0)$.

First, we obtain the model characteristic equations, which are ODEs valid along the characteristic curves, and use them to write a Taylor-series expansion of the particle density in each state about the initial time. This procedure is systematic: if the PDE input-output equation is order $k$, then $k$ Cauchy data are required to fix the solution for the total particle density according to the input-output equation. The total particle density, $N(x,t)$, needs to expanded about $t=0$ with Taylor-series expansions to the order $k-1$. Then, by evaluating at time $t=0$ the expansion for $N(x,t)$ and its derivatives up to order $k-1$, we obtain the $k$ initial-condition input-output equations corresponding to the required Cauchy data. We note that fixing all time derivatives of the total particle density at $t=0$ (required up to the order in time of the input-output PDE minus one) fixes the time evolution of the total particle density.

The Taylor-series expansion of the total particle density at the initial time requires the existence of the characteristic equations, which is guaranteed in our case since the PDE system is \textit{strongly hyperbolic}~\citep{godlewski1991hyperbolic}. Next, we write the characteristic curves of the PDE system as $$\frac{\text{d} x}{\text{d} t}=v_s,$$
for $s=1,2,3$. For each characteristic $\text{d} x/\text{d} t=v_s$, denote the position of a particle along the characteristic $x_s(t)$ and its initial location $\xi_s=x_s(t=0)$. Then, we write the characteristic equations as follows
\begin{equation}\label{Eq:ODE}
    D_sn_s(x,t):=\left(\partial_t
    +v_s\partial_x \right)n_s(x(t),t)=\sum_{u=1}^3 q_{us}n_u(x(t),t), \ \text{ on } x(t)=\xi_s+v_st,
\end{equation}
for $s=1,2,3$.

Next, we Taylor expand the density in state $s$ along its characteristic about
$t=0$. Along the characteristic $x_s(t)=\xi_s+v_st$, the second-order
Taylor expansion is
\begin{equation}\label{Eq:Taylor_ns}
\begin{aligned}
n_s(\xi_s+v_st,t)
={}&n_s(\xi_s,0)
+tD_sn_s(\xi_s,0)
+\frac{t^2}{2}D_s^2n_s(\xi_s,0)
+O(t^3),
\end{aligned}
\end{equation}
for $s=1,2,3$. From Equation~\eqref{Eq:ODE},
$$D_s^2 n_s=\sum_{u=1}^3q_{us}D_s n_u.$$
Using the PDE for $n_u$ gives
$$\begin{aligned}
D_sn_u=\partial_t n_u
+v_s\partial_x n_u=
\left(
\partial_t n_u
+v_u\partial_x n_u
\right)
+(v_s-v_u)\partial_x n_u=
\sum_{r=1}^3q_{ru}n_r
+(v_s-v_u)\partial_x n_u.
\end{aligned}$$
Therefore,
\begin{equation}\label{Eq:Ds2ns}
D_s^2n_s
=
\sum_{u=1}^3\sum_{r=1}^3q_{us}q_{ru}n_r
+
\sum_{u=1}^3q_{us}(v_s-v_u)
\partial_x n_u.
\end{equation}

Using the initial condition $n_s(x,0)=a_sf(x)$, so that
$\partial_xn_s(x,0)=a_sf'(x)$, Equations~\eqref{Eq:Taylor_ns}
and~\eqref{Eq:Ds2ns} give
\begin{equation}\label{Eq:ODE_0}
\begin{aligned}
n_s(\xi_s+v_st,t)
={}&a_sf(\xi_s)
+t\left(\sum_{u=1}^3q_{us}a_u\right)f(\xi_s)\\
&+\frac{t^2}{2}
\left(\sum_{u=1}^3\sum_{r=1}^3q_{us}q_{ru}a_r\right)
f(\xi_s)\\
&+\frac{t^2}{2}
\left(\sum_{u=1}^3q_{us}(v_s-v_u)a_u\right)
f'(\xi_s)
+O(t^3),
\end{aligned}
\end{equation}
for all $\xi_s\in\mathbb{R}$ and $s=1,2,3$. To obtain the expansion of the total density at a fixed location
$\xi\in\mathbb{R}$, we choose $\xi_s=\xi-v_st$ and sum
Equation~\eqref{Eq:ODE_0} over $s$. This gives
\begin{equation}\label{Eq:N(xi,t)}
\begin{aligned}
N(\xi,t)
={}&
\sum_{s=1}^3\left(
a_s
+t\sum_{u=1}^3q_{us}a_u
+\frac{t^2}{2}
\sum_{u=1}^3\sum_{r=1}^3q_{us}q_{ru}a_r
\right)
f(\xi-v_st)\\
&+
\frac{t^2}{2}
\sum_{s=1}^3\sum_{u=1}^3
q_{us}(v_s-v_u)a_u
f'(\xi-v_st)
+O(t^3).
\end{aligned}
\end{equation}
Next, we Taylor expand
\begin{equation}
f(\xi-v_st)
=
f(\xi)-v_stf'(\xi)
+\frac{v_s^2t^2}{2}f''(\xi)
+O(t^3),
\end{equation}
and
\begin{equation}
f'(\xi-v_st)=f'(\xi)+O(t),
\end{equation}
and, by substitution into Equation~\eqref{Eq:N(xi,t)} and rearranging, we obtain
\begin{equation}
\label{Eq:N_expanded}
\begin{aligned}
N(\xi,t)
={}&
\left(\sum_{s=1}^3a_s\right)f(\xi)
\\
&+
t\left[
\left(\sum_{s=1}^3\sum_{u=1}^3q_{us}a_u\right)f(\xi)
-
\left(\sum_{s=1}^3a_sv_s\right)f'(\xi)
\right]
\\
&+
\frac{t^2}{2}
\Bigg[
\left(\sum_{s=1}^3a_sv_s^2\right)f''(\xi)
\\
&\qquad \quad +
\left(
\sum_{s=1}^3\sum_{u=1}^3
q_{us}(v_s-v_u)a_u
-
2
\sum_{s=1}^3v_s\sum_{u=1}^3q_{us}a_u
\right)f'(\xi)
\\
&
\qquad \quad + \left(
\sum_{s=1}^3\sum_{u=1}^3\sum_{r=1}^3
q_{us}q_{ru}a_r
\right)f(\xi)
\Bigg]
\\
&
+O(t^3).
\end{aligned}
\end{equation}
Since, by definition, $a_1+a_2+a_3=1$ and $q_{u1}+q_{u2}+q_{us}=0$ for $u=1,2,3$, expanding and simplifying Equation~\eqref{Eq:N_expanded}, we obtain
\begin{equation}\label{Eq:N_c9-c11}
N(\xi,t)
=
f(\xi)
-c_9t f'(\xi)
+
\frac{t^2}{2}
\left[
c_{10}f''(\xi)-c_{11}f'(\xi)
\right]
+O(t^3),
\end{equation}
where we define
\begin{subequations}\label{Eq:c10-c11}
\begin{align}
c_{10}
&=
\sum_{s=1}^3a_sv_s^2,\label{Eq:c10}
\\
c_{11}
&=
\sum_{s=1}^3v_s\sum_{u=1}^3q_{us}a_u,\label{Eq:c11}
\end{align}
\end{subequations}
and $c_9$ is as defined in Equation~\eqref{Eq:c9_IC}. We also note that $c_{11}$ can be written in terms of the rates and probability parameters as
\begin{equation}\label{Eq:c11b}
\begin{aligned}
c_{11}
={}&
\lambda_1a_1
\left(
v_3-v_1+(v_2-v_3)p_{12}
\right)
\\
&+
\lambda_2a_2
\left(
v_3-v_2+(v_1-v_3)p_{21}
\right)
\\
&+
\lambda_3(1-a_1-a_2)
\left(
v_2-v_3+(v_1-v_2)p_{31}
\right).
\end{aligned}
\end{equation}

Evaluating Equation~\eqref{Eq:N_c9-c11} and its derivatives in time at $t=0$, and setting $\xi=x$ we obtain the following initial-condition input-output equations
\begin{gather}
    N(x,0) = f(x),\label{Eq:Tay_IO_1}
    \\
    \partial_t N (x,0) = -c_9 f'(x),
    \label{Eq:Tay_IO_2}
    \\
    \partial_{tt} N(x,0)= c_{10}f''(x)-c_{11}f'(x).
    \label{Eq:Tay_IO_3}
\end{gather}
We note that Equations~\eqref{Eq:Tay_IO_1}-\eqref{Eq:Tay_IO_2} are the same as the ones obtained in Equations~\eqref{Eq:RA_IC_IO_base0}-\eqref{Eq:IC_IOeq}, and only give the identifiability of $c_9$. Moreover, Equation~\eqref{Eq:Tay_IO_3} provides an expression for $\partial_{tt} N(x,0)$, that is the third Cauchy datum, required to fix the solution of the third-order input-output equation in Equation~\eqref{Eq:IO}. Equation~\eqref{Eq:Tay_IO_3}, since $f'$ and $f''$ are linearly independent, gives identifiability of $c_{10}$ and $c_{11}$. Hence, $c_1,\ldots,c_{11}$ are all the structurally identifiable parameter combinations.

\subsection{A Fourier-transform approach for the derivation of the initial-condition input-output equations}
\label{Subsec:Fourier_IC}

In this section, we propose a Fourier-transform approach to derive the input-output equation for $\partial_{tt}N(x,0)$ associated with the specified separable initial condition. More generally, we show that this approach allows systematic derivation of all the initial-condition input-output equations for the model (Equations~\eqref{Eq:Tay_IO_1}-\eqref{Eq:Tay_IO_3}). The choice between the Taylor-series-expansion or the Fourier-transform approaches depends on their applicability to the mathematical structure of the PDE model considered. Applying the Fourier transform in space converts spatial derivatives in the PDE system into algebraic expressions, resulting in a system of ODEs in time. The required Cauchy data can then be obtained systematically by successively differentiating the transformed total particle density with respect to time and evaluating these derivatives at $t=0$ using the specified initial condition. As in Section~\ref{Subsec:Taylor_IC}, we note that fixing all time derivatives of the total particle density at $t=0$ (required up to the order in time of the input-output PDE minus one) fixes the time evolution of the total particle density. Finally, taking the inverse Fourier transform gives the corresponding initial-condition input-output equations in the spatial domain.

We define the Fourier transform in the spatial variable as
\begin{equation*}
    \widehat{g}(k)=\int_{\mathbb{R}}g(x)\exp(-\mathrm{i}kx)\,\mathrm{d}x,
\end{equation*}
where $\mathrm{i}$ is the imaginary unit, and its inverse as
\begin{equation*}
    g(x)=\frac{1}{2\pi}\int_{\mathbb{R}}\widehat{g}(k)\exp(\mathrm{i}kx)\,\mathrm{d}k,
\end{equation*}
with the convention, 
\begin{equation*}
    \widehat{\frac{\partial^c g}{\partial x^c}}=(\mathrm{i}k)^c\widehat{g}, \qquad c\in \mathbb{N}.
\end{equation*}

Applying the Fourier transform in space to the model in Equation~\eqref{Eq:RA} gives
\begin{equation}\label{Eq:Fourier_ns_t}
    \partial_t\widehat{n}_s(k,t)=\sum_{u=1}^{3}q_{us}\widehat{n}_u(k,t)-\mathrm{i}kv_s\widehat{n}_s(k,t),
\end{equation}
for $s=1,2,3$, and the Fourier transform of the total particle density is
\begin{equation}\label{Eq:Fourier_eta}
    \widehat{N}(k,t)=\sum_{s=1}^{3}\widehat{n}_s(k,t).
\end{equation}
Moreover, the initial condition in Equation~\eqref{Eq:IC} gives
\begin{equation}\label{Eq:Fourier_IC}
    \widehat{n}_s(k,0)=a_s\widehat{f}(k),\qquad s=1,2,3.
\end{equation}
First, using Equation~\eqref{Eq:Fourier_IC} and taking the inverse Fourier transform of Equation~\eqref{Eq:Fourier_eta} gives
\begin{equation}\label{Eq:Fourier_IO_1}
    N(x,0) = f(x),
\end{equation}
which recovers the first initial-condition input-output equation in Equation~\eqref{Eq:Tay_IO_1}.

Next, we recover the second initial-condition input-output equation in Equation~\eqref{Eq:Tay_IO_2}. Differentiating Equation~\eqref{Eq:Fourier_eta} with respect to time and substituting with Equation~\eqref{Eq:Fourier_ns_t}, we obtain
\begin{equation}\label{Eq:N_t^hat}
    \partial_t\widehat{N}(k,t)=\sum_{s=1}^{3}\sum_{u=1}^{3}q_{us}\widehat{n}_u(k,t)-\mathrm{i}k\sum_{s=1}^{3}v_s\widehat{n}_s(k,t).
\end{equation}
Since the rows of the transition matrix sum to zero, Equation~\eqref{Eq:N_t^hat} becomes
\begin{equation*}
    \partial_t\widehat{N}(k,t)=-\mathrm{i}k\sum_{s=1}^{3}v_s\widehat{n}_s(k,t),
\end{equation*}
which, evaluated at $t=0$ and using Equation~\eqref{Eq:Fourier_IC}, gives
\begin{equation*}
    \partial_t\widehat{N}(k,0)=-\mathrm{i}k\sum_{s=1}^{3}a_sv_s\widehat{f}(k)=-c_9\mathrm{i}k\widehat{f}(k),
\end{equation*}
with $c_9$ defined in Equation~\eqref{Eq:c9_IC}. Using $\widehat{f'}(k)=\mathrm{i}k\widehat{f}(k)$, taking the inverse Fourier transform, and substituting $\xi=x$ gives
\begin{equation}\label{Eq:Fourier_IO_2}
    \partial_tN(x,0)=-c_9f'(x).
\end{equation}

Finally, we recover the third initial-condition input-output equation in Equation~\eqref{Eq:Tay_IO_3}. We differentiate Equation~\eqref{Eq:Fourier_eta} twice with respect to time to obtain
\begin{equation*}
    \partial_{tt}\widehat{N}(k,t)=-\mathrm{i}k\sum_{s=1}^{3}v_s\partial_{t}\widehat{n}_s(k,t).
\end{equation*}
Substituting Equation~\eqref{Eq:Fourier_ns_t} gives
\begin{equation*}
    \partial_{tt}\widehat{N}(k,t)=-\mathrm{i}k\sum_{s=1}^{3}\sum_{u=1}^{3}v_sq_{us}\widehat{n}_u(k,t)-k^2\sum_{s=1}^{3}v_s^2\widehat{n}_s(k,t),
\end{equation*}
which, evaluated at $t=0$ and using Equation~\eqref{Eq:Fourier_IC}, gives
\begin{equation*}
    \partial_{tt}\widehat{N}(k,0)=-\mathrm{i}k\left(\sum_{s=1}^{3}\sum_{u=1}^{3}v_sq_{us}a_u\right)\widehat{f}(k)-k^2\left(\sum_{s=1}^{3}a_sv_s^2\right)\widehat{f}(k).
\end{equation*}
From Equation~\eqref{Eq:c10-c11}, the coefficients $c_{10}$ and $c_{11}$ are
\begin{equation*}
    c_{10}=\sum_{s=1}^{3}a_sv_s^2,\qquad c_{11}=\sum_{s=1}^{3}v_s\sum_{u=1}^{3}q_{us}a_u=\sum_{s=1}^{3}\sum_{u=1}^{3}v_sq_{us}a_u.
\end{equation*}
Therefore,
\begin{equation*}
    \partial_{tt}\widehat{N}(k,0)=\left(-c_{10}k^2-c_{11}\mathrm{i}k\right)\widehat{f}(k).
\end{equation*}
Using $\widehat{f''}(k)=-k^2\widehat{f}(k)$ and $\widehat{f'}(k)=\mathrm{i}k\widehat{f}(k)$, taking the inverse Fourier transform, and substituting $\xi=x$, gives
\begin{equation}\label{Eq:Fourier_IO_3}
    \partial_{tt}N(x,0)=c_{10}f''(x)-c_{11}f'(x).
\end{equation}

Equations~\eqref{Eq:Fourier_IO_1}-\eqref{Eq:Fourier_IO_2} are the same as the ones obtained in Equations~\eqref{Eq:RA_IC_IO_base0}-\eqref{Eq:IC_IOeq}, and only give the identifiability of $c_9$. Moreover, Equation~\eqref{Eq:Fourier_IO_3} provides an expression for $\partial_{tt} N(x,0)$, that is the third Cauchy datum, required to fix the solution of the third-order input-output equation in Equation~\eqref{Eq:IO}. Equation~\eqref{Eq:Fourier_IO_3}, since $f'$ and $f''$ are linearly independent, gives identifiability of $c_{10}$ and $c_{11}$. Hence, $c_1,\ldots,c_{11}$ are all the structurally identifiable parameter combinations. As expected, the Taylor-series-expansion and Fourier-transform approaches yield the same initial-condition input-output equations (Equations~\eqref{Eq:Fourier_IO_1}-\eqref{Eq:Fourier_IO_3} and Equations~\eqref{Eq:Tay_IO_1}-\eqref{Eq:Tay_IO_3}), since they provide two alternative ways to obtain expressions of the total particle density and its derivatives at the initial time. We now combine the identifiable coefficients $c_9,c_{10},c_{11}$ with $c_1,\ldots,c_8$ from the input-output equation in Equation~\eqref{Eq:IO} to complete the structural identifiability analysis of the model.

\subsection{The model with separable initial condition is locally structurally identifiable measuring the total particle density}

Now, we analyse the coefficients $c_1,\ldots,c_{11}$, obtained in Sections~\ref{Subsec:DA}-\ref{Subsec:Fourier_IC}, to conclude the structural identifiability analysis. In Section~\ref{Subsec:DA}, we obtained that, after fixing a state labelling from the coefficients $c_1,\ldots,c_6$, the velocities $v_1,v_2,v_3$ and switching rates $\lambda_1,\lambda_2,\lambda_3$ are structurally identifiable (Equation~\eqref{Eq:sys_lambda}). In addition, the equation $a_1+a_2+a_3=1$ and the coefficients $c_9,c_{10}$ form a Vandermonde system in $[a_1,a_2,a_3]^T$~\citep{horn2012matrix}. Therefore, the initial condition parameters $a_1,a_2,a_3$ are also structurally identifiable. Finally, we are left with the analysis of the coefficients $c_7,c_8,c_{11}$ to determine whether the probability parameters $p_{12},p_{21},p_{31}$ are structurally identifiable.

\begin{figure*}[!ht]
    \centering
    \includegraphics[width=\textwidth]{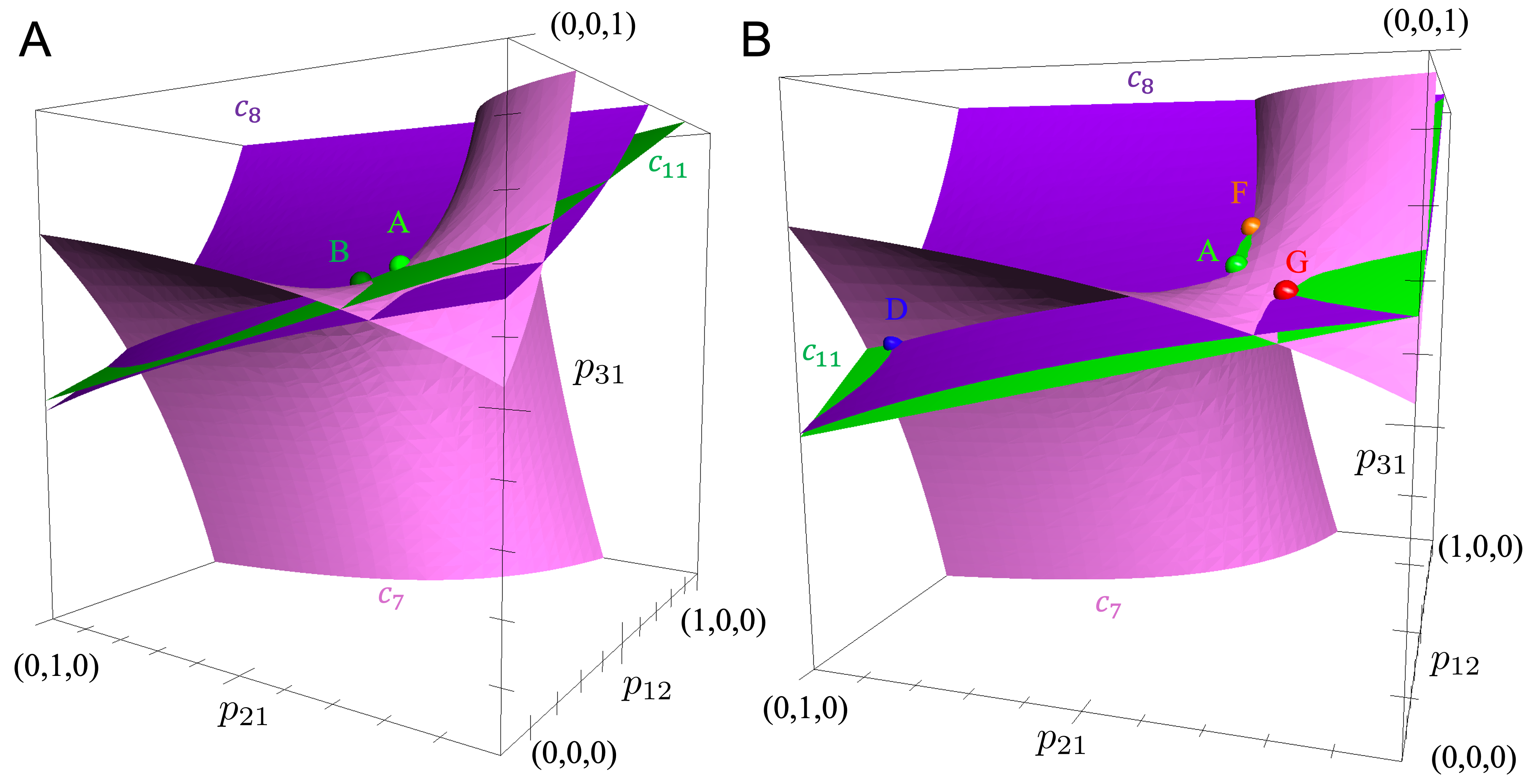}
    \caption[Intersection between the surfaces given by coefficients $c_7,c_8$ and the plane $c_{11}$]{\textbf{Intersection between the surfaces given by coefficients $c_7,c_8$ and the plane $c_{11}$ for some example parameter sets.} \textbf{A} is generated using the parameter set $\boldsymbol{\theta}_A=\{v_1=20, v_2=-15, v_3=0, \lambda_1=1, \lambda_2=1/2, \lambda_3=3/10, p_{12}=1/5, p_{21}=3/10, p_{31}=7/10\}$ and initial condition parameters $\boldsymbol{a}_A=\boldsymbol{\pi}_A=\{a_1=237/1441, a_2=24/131, a_3=940/1441\}$. The three coefficients, $c_7,c_8,c_{11}$, intersect at the point $\text{A}=\{p_{12}=1/5, p_{21}=3/10, p_{31}=7/10\}$ and at the point $\text{B}=\{p_{12}=66/395, p_{21}=79/220, p_{31}=158/235\}$. \textbf{B} is generated using the parameter set $\boldsymbol{\theta}_A$ as in \textbf{A}, but with initial condition $\boldsymbol{a}_A=\{a_1=1/4, a_2=3/20, a_3=3/5\}$. Now, the intersection points are $\text{A}=\{p_{12}=1/5, p_{21}=3/10, p_{31}=7/10\}$, $\text{D}=\{p_{12}=1/5, p_{21}=47/50, p_{31}=23/42\}$, $\text{F}=\{p_{12}=(405+3\sqrt{4001})/2000, p_{21}=(997+3\sqrt{4001})/4000, p_{31}=(999+\sqrt{4001})/1400\}$, and $\text{G}=\{p_{12}=(405-3\sqrt{4001})/2000,p_{21}=(997-3\sqrt{4001})/4000,p_{31}=(999-\sqrt{4001})/1400\}$. These images were created using Apple Grapher (macOS).}
    \label{Fig:3}
\end{figure*}

The parameters $p_{12},p_{21},p_{31}$ must lie in the intersection between the plane defined by $c_{11}$ (Figure~\ref{Fig:3}, green planes) and the quadratic surfaces defined by $c_7$ and $c_8$ (Figure~\ref{Fig:3}, pink and purple surfaces). By Bézout's theorem, there are at most four isolated solutions counted with multiplicity~\citep{fulton1969algebraic}, therefore the probability parameters $p_{12},p_{21},p_{31}$ are, in general, only locally structurally identifiable. However, the number of physically admissible solutions in $[0,1]^3$ may be smaller. The probability parameters become globally structurally identifiable when only one solution of the intersection between $c_7,c_8,c_{11}$ lies within the admissible parameter domain $[0,1]^3$. The number of admissible solutions for known values of $c_7,c_8,c_{11}$ can be determined by solving the system and selecting the solutions in $[0,1]^3$ (for example, see the Mathematica notebook \textit{Section\_4.4\_locally\_SI\_coefficients.nb}). Next, we show that the number of admissible solution varies with the initial condition parameters.

Figure~\ref{Fig:3}\textbf{A} shows that, for an example parameter set $\boldsymbol{\theta}_A$ (A) with initial condition $\boldsymbol{a}_A=\boldsymbol{\pi}_A$, there is only one other admissible probability parameter set (B) with the same coefficients $c_1,\ldots,c_{11}$. Figure~\ref{Fig:3}\textbf{B} shows that, for the example parameter set $\boldsymbol{\theta}_A$ (A) with initial condition $\boldsymbol{a}_A=[1/4, 3/20, 3/5]$, there are three other probability parameter sets (D,F,G) with the same coefficients $c_1,\ldots,c_{11}$. In Supplementary Information Section~S5.2, we show the numerical solutions of the examples in Figure~\ref{Fig:3}\textbf{A} and~\textbf{B}. Moreover, we show other examples of parameter sets equivalent to $\boldsymbol{\theta}_A$, as the initial condition $\boldsymbol{a}_A$ is varied (also see the Mathematica notebook \textit{Section\_4.4\_locally\_SI\_coefficients.nb}).

We recall that, in Section~\ref{Sec:SI_SPT}, we found that the model parameters are structurally identifiable measuring infinitely many, infinitely long single-particle trajectories. In contrast, with total particle density data, the model with separable initial condition is, in general, only locally structurally identifiable, since up to four sets of transition probabilities may give the same observable output. Hence, single-particle trajectory and total particle density data can lead to different structural identifiability properties.

In Supplementary Information Sections~S4 and~S6, we analyse the case with two equal velocities, $v_2=v_3=v$, and we obtain that $v_1,v,\lambda_1,a_1$ and $k_1,k_2,k_3,k_4$ (Equations~\eqref{Eq:k_1k_2k_3}-\eqref{Eq:k_4}) are identifiable. In this case, the structural identifiability properties with total particle density data are the same as those obtained given infinitely many, infinitely long single-particle trajectories, obtained in Supplementary Information Section~S3. Indeed, when states $s=2$ and $s=3$ share the same velocity, they are indistinguishable and can be viewed as a single effective state with velocity $v$. The model can therefore be interpreted, at the level of the observations, as a two-velocity-state process, with velocities $v_1$ and $v$, but with a non-exponential waiting-time distribution in the state with velocity $v$. The parameters governing this waiting-time distribution ($\lambda_2,\lambda_3,p_{12},p_{21},p_{31},a_2$) are non-identifiable and can only be recovered through the combinations $k_1,k_2,k_3,k_4$. Consequently, both data types identify the velocity $v$, the parameters uniquely associated with state $s=1$ ($v_1,\lambda_1,a_1$), and the parameter combinations $k_1,k_2,k_3,k_4$.

We highlight that, in Section~\ref{Sec:SI_SPT}, we found that the model parameters are structurally identifiable measuring single-particle trajectories, while here we find that they are only locally identifiable measuring the total particle density evolution. Hence, single-particle trajectory and total particle density data lead to different structural identifiability properties.

\section{Discussion and conclusions}\label{Sec:discussion}

In this paper, we developed a general methodology for the analysis of the structural identifiability of a class of spatio-temporal stochastic processes, for which no broadly applicable techniques currently exist. We demonstrated our methodology by applying it to a velocity-jump process in which the particle's motion is determined by its hidden internal state, which evolves within a network of three states, with each state characterised by a constant velocity and constant switching rates. The particles' initial locations are sampled according to a probability distribution function $f(x)$, and their initial states are sampled according to a probability vector $[a_1,a_2,a_3]^T$. 

We first introduced and applied a method to study the identifiability properties of partially-observed Markov models given single-particle trajectory data. Second, we investigated the structural identifiability of stochastic models with total particle density data by formulating a PDE description. We applied the differential algebra approach to reduce the PDE system to input-output equations whose coefficients determine the set of identifiable parameter combinations. We also analysed the initial condition, which is known to influence structural identifiability~\citep{chis2011structural,saccomani2003parameter,ljung1994global,diop1991nonlinear,tunali1987new}, as it may contain information about the model parameters that is not captured by the PDE input-output equations.

An important question in the structural identifiability analysis of PDE models is how to systematically incorporate the information arising from specified initial conditions. We considered separable initial conditions $n_s(x,0)=a_sf(x)$, for $s=1,2,3$, such that $a_1+a_2+a_3=1$. Since the PDE input-output equation in Equation~\eqref{Eq:IO} is third order in time, three initial-time Cauchy data guarantee a unique solution for the total particle density. Specifically, we showed that the initial-condition input-output equations for $N(x,0)$ and $\partial_t N(x,0)$, obtained with existing approaches~\citep{browning2024structural,renardy2022structural}, are not sufficient to capture all identifiable parameter combinations. Hence, the initial-condition input-output equation for $\partial_{tt} N(x,0)$ is required to complete the structural identifiability analysis.

We proposed two approaches to systematically derive all initial-condition input-output equations for the class of velocity-jump PDE models considered here. Specifically, we developed Taylor-series-expansion and Fourier-transform approaches, both of which use ODE formulations of the total particle density to derive the initial-condition input-output equations by successive time differentiation and evaluation at the initial time. These approaches therefore provide two complementary and systematic ways to incorporate the specified initial conditions into structural identifiability analysis, although their applicability may depend on the mathematical setting.

On the one hand, the Taylor-series-expansion approach extends ideas previously used in ODE structural identifiability~\cite{pohjanpalo1978system} and is applicable to systems that admit characteristic equations, such as the velocity-jump models considered here and other first-order hyperbolic systems. On the other hand, the Fourier-transform approach derives input-output equations through successive differentiation of the transformed total particle density. We demonstrated this approach on an unbounded spatial domain with far-field boundary conditions, where the Fourier transform converts spatial derivatives into algebraic expressions. A natural extension is to periodic bounded domains, where Fourier series provide an analogous representation of the transformed ODE system. For other boundary conditions, the transformation of spatial derivatives may introduce additional boundary terms, and accounting for boundary conditions in structural identifiability analysis provides a direction for future work.

The initial-condition input-output equations recovered by the Taylor-series-expansion and Fourier-transform approaches give the additional identifiable parameter combinations $c_9,c_{10},c_{11}$. We found that the three-state model considered, with distinct velocities, is structurally identifiable given infinitely many, infinitely long single-particle trajectories. In contrast, with total particle density data, the model is not structurally identifiable for any unknown form of initial condition, but it is locally structurally identifiable assuming the separable form of initial condition specified in Equation~\eqref{Eq:IC}. Specifically, from total particle density data, the velocities, switching rates, and initial condition parameters are globally structurally identifiable once a state labelling is fixed, whereas up to four sets of switching probability parameters may produce the same observable output.

We also consider the case in which two states share the same velocity since single-particle trajectory data are not sufficient to reveal the internal state. In Supplementary Information Section~S3, we provided a novel method to study structural identifiability given single-particle trajectory data using the waiting time distribution for states with the same observed velocity, computed using recurrence relations. Moreover, in Supplementary Information Sections~S4 and~S6, we studied the identifiability properties of the model measuring the total particle density when two states share the same velocities.

If $v_2=v_3=v$, single-particle trajectory and total particle density data lead to the same structural non-identifiability. Indeed, states $s=2$ and $s=3$ are indistinguishable and can be viewed as a single effective state with velocity $v$, with a non-exponential waiting-time distribution determined by the underlying state-switching dynamics. Consequently, both observation schemes identify $v_1$, $v$, $\lambda_1$, $a_1$, and the same parameter combinations $k_1,k_2,k_3,k_4$ (Equations~\eqref{Eq:k_1k_2k_3} and~\eqref{Eq:k_4}), but cannot identify the parameters governing the two shared-velocity states individually.

We also showed that the sets of structurally equivalent parameter values to a specified model depend on the initial condition parameters. For example, we showed that the parameter sets $\boldsymbol{\theta}_A$ and $\boldsymbol{\theta}_B$ produce the same model solutions under a specific initial condition (Figure~\ref{Fig:3}\textbf{A}), but not for a different initial condition (Figure~\ref{Fig:3}\textbf{B}). This highlights the connection between experimental design and structural identifiability and suggests that, when experimentally feasible, varying the initial condition may provide additional information for distinguishing between otherwise equivalent parameter sets. 

We highlight that our structural identifiability analysis from single-particle trajectory data assumes infinitely many, infinitely long, continuously observed trajectories. Structural identifiability is a necessary, but not sufficient, condition for practical identifiability, which concerns the precision with which parameters can be estimated from available data. In practice, several factors may affect parameter estimation. A key challenge arises from discrete-time sampling, since changes in velocity may occur between observation times, preventing the direct observation of the underlying switching events~\citep{ceccarelli2026}. Moreover, the finite number and duration of the observed trajectories and the measurement noise also affect the uncertainty of parameter estimates~\citep{ceccarelli2026}.

Several avenues for future work emerge from this study. Although we demonstrated the methodology using a three-state model, it applies equally to models with an arbitrary number of states, as specified in~\citep{ceccarelli2026,ceccarelli2025}. For single-particle trajectory data, the identifiability results obtained in Section~\ref{Sec:SI_SPT} can be trivially extended to the model with any number of states. In particular, the number of states can be identified as well as all model parameters as long as the velocities are all distinct. For the total particle density data, although the methodology is systematic and, in principle, broadly applicable, we expect the computational complexity to remove unmeasured quantities to grow rapidly, as the number and order of derivatives required typically increase with the number of hidden states.

Instead of observing the total particle density data, experimental techniques may only be able to detect stationary or slow-moving particles~\citep{syga2018method}. Therefore, idealised data could be assumed to only observe particles in a specific state, $n_s(x,t)$, for example, with zero velocity. The methods we proposed are also directly applicable to study the identifiability properties of models when the data capture the particle density in a specific state or measuring any linear combination of the quantities in each state. We expect these data to lead to analogous identifiability properties of the model as long as the linear combination coefficients are known.

Finally, the two types of data we consider, that capture either individual behaviour or particle density evolution, can be described by two natural mathematical representations, also typical of chemical reaction networks~\citep{anderson2011continuous,kurtz1980representations}: particle-based models and differential-equation-based models. Hence, future work may be done to extend the methodology to chemical reaction models given single-particle trajectory and total particle density data, particularly with relevant spatial components.

Overall, this work introduces novel methods to analyse the structural identifiability of a class of spatio-temporal stochastic models, including Taylor-series-expansion and Fourier-transform approaches for incorporating initial conditions into the analysis. We show how structural identifiability depends on the type of data available, namely single-particle trajectory data or total particle density data. Finally, we identify potential calibration challenges arising from the lack of structural identifiability and propose practical strategies, based on experimentally controlling the initial conditions, to recover parameter identifiability.

%\bmhead*{Acknowledgements}

\section*{Declarations}

\bmhead*{Funding}
A.C. is supported by a Mathematical Institute Studentship from the University of Oxford. R.E.B. is supported by a grant from the Simons Foundation (MP-SIP-00001828). For the purpose of open access, the author has applied a CC BY public copyright licence to any author accepted manuscript arising from this submission.

\bmhead*{Conflict of interest}
The authors declare that they have no conflict of interest.

%\bmhead{Ethics approval and consent to participate}?

\bmhead*{Consent for publication}
All the authors approved the final version of the manuscript.

\bmhead*{Code availability}
The Python files, Mathematica notebook and Apple Grapher files are available in the GitHub repository \href{https://github.com/a-ceccarelli/structural_identifiability_stochastic_processes}{a-ceccarelli/structural\_identifiability\_stochastic\_processes} and in the Zenodo repository \href{https://doi.org/10.5281/zenodo.23688894}{10.5281/zenodo.23688894}.

%\bmhead*{Data availability}

%\bmhead*{Materials availability}
%Not applicable.

\bmhead*{CRediT author statement}
A.C.: Writing - Original Draft, Writing - Review and Editing, Conceptualisation, Methodology, Formal Analysis, Visualisation, Software Implementation. A.P.B. and R.E.B.: Supervision, Conceptualisation, Writing - Review and Editing.

\singlespacing

%\bibliographystyle{plainnat} 
% Bibliography name
%\bibliography{}
%\linespread{0.95}
\selectfont\bibliography{sn-bibliography}

\end{document}

\title{\centering \textbf{Supplementary Information} \\ Structural identifiability of partially-observed \\ stochastic processes: from single-particle \\ trajectories to total particle density data}

\author*[1]{\fnm{Arianna} \sur{Ceccarelli}\orcidlink{0000-0002-9598-8845}}\email{ceccarelli@maths.ox.ac.uk}

\author[2]{\fnm{Alexander P.} \sur{Browning}\orcidlink{0000-0002-8753-1538}}

\author[1]{\fnm{Ruth E.} \sur{Baker}\orcidlink{0000-0002-6304-9333}}

\affil[1]{\centering
Mathematical Institute, University of Oxford, UK}%\\Woodstock Road, Oxford, OX2 6GG
\affil[2]{\centering School of Mathematics and Statistics, University of Melbourne, Australia}

\renewcommand\theequation{S\arabic{equation}}
\renewcommand\thefigure{S\arabic{figure}}
\renewcommand\thesection{S\arabic{section}}
\renewcommand\thetable{S\arabic{table}}

\maketitle

\tableofcontents

\section{The stochastic model stationary distribution}\label{Sec:S1}

We consider the three-state velocity-jump process introduced in Section~2 in the main text. The state evolves as a continuous-time Markov chain, which, if irreducible, has a unique stationary distribution (see~\citep{ceccarelli2026,ceccarelli2025}). We can compute the stationary distribution of the stochastic process in terms of the parameters of the transition matrix $\boldsymbol{Q}$. If $\boldsymbol{\pi}=[\pi_1,\pi_2,\pi_3]^T$ is any eigenvector in the kernel of $\boldsymbol{Q}^T$, such that $\boldsymbol{Q}^T \boldsymbol{\pi}=\boldsymbol{0}$, then the stationary distribution $\boldsymbol{w}=[w_1,w_2,w_3]^T$ is the unique solution of the system $\boldsymbol{Q}^T \boldsymbol{w}=\boldsymbol{0}$ with $w_s\ge0$ for $s=1,2,3$ and $w_3=1-w_1-w_2$, obtained as $\boldsymbol{w}=\boldsymbol{\pi}/(\pi_1+\pi_2+\pi_3)$. 

The stationary distribution $\boldsymbol{w}$ is unique and can be written in terms of the switching rates and probability parameters as
\begin{equation}\label{Eq:steady_three-state}
\boldsymbol{w}=\frac{[\phi_1, \phi_2, \phi_3]^T} {\phi_1+\phi_2+\phi_3}, \quad \text{with} \quad \phi_s=\lambda_u\lambda_z(1-p_{uz}p_{zu}), 
\end{equation}
for $(s,z,u)\in\{(1,2,3), (2,1,3), (3,1,2)\}$.
Equivalently, the stationary distribution can be written in terms of the switching rates and the probability parameters $p_{12},p_{21},p_{31}$ as
\begin{equation}\label{Eq:steady_three-state_explicit}
\begin{aligned}
    &w_1= \frac{\lambda_2 \lambda_3 (1- (1 - p_{21})(1 - p_{31})) }{\lambda_2 \lambda_3 (1- (1 - p_{21})(1 - p_{31})) + \lambda_1 \lambda_3 (1 - (1 - p_{12}) p_{31}) + \lambda_1 \lambda_2 (1 - p_{12} p_{21})},
    \\
    &w_2= \frac{\lambda_1 \lambda_3 (1 - (1 - p_{12}) p_{31})}{\lambda_2 \lambda_3 (1- (1 - p_{21})(1 - p_{31})) + \lambda_1 \lambda_3 (1 - (1 - p_{12}) p_{31}) + \lambda_1 \lambda_2 (1 - p_{12} p_{21})},
    \\
    &w_3= \frac{\lambda_1 \lambda_2 (1 - p_{12} p_{21})}{\lambda_2 \lambda_3 (1- (1 - p_{21})(1 - p_{31})) + \lambda_1 \lambda_3 (1 - (1 - p_{12}) p_{31}) + \lambda_1 \lambda_2 (1 - p_{12} p_{21})}.
\end{aligned}
\end{equation}

\section{The PDE model derivation}

In this section, we derive the PDE system describing the evolution of the particle density in each state $s=1,2,3$ at location $x$ and at time $t$, $n_s(x,t)$, as a Fokker-Planck equation. 

We fix a small time increment $\Delta t$, and denote the density in state $s$, at location $x$ and time $t+\Delta t$, $n_s(x,t+\Delta t)$, which we write as a sum of terms up to order $o(\Delta t)$. Moreover, as $\Delta t$ can be chosen arbitrarily small, particles may only switch state once in the time interval $[t,t+\Delta t)$, as a number of switches greater than one would have order $o(\Delta t)$.

The density of particles in state $s$ at time $t$, at the location $x-v_s\Delta t$ that remain in state $s$ can be written as $(1-\lambda_s\Delta t)n_s(x-v_s\Delta t,t)$. Then, we add, for each state $u\ne s$, the density of particles at state $u$, that switch state exactly once, to state $s$, at time $t+\hat\Delta t\in[t, t+\Delta t)$, with $0\le\hat\Delta t<\Delta t$, which at time $t$ is at location $x-v_u\hat\Delta t-v_s(\Delta t-\hat\Delta t)$, and can be written as
$(\lambda_up_{us}\Delta t) n_u(x-v_u\hat\Delta t-v_s(\Delta t-\hat\Delta t),t)$. Hence, we obtain
$$n_s(x,t+\Delta t)=(1-\lambda_s\Delta t)n_s(x-v_s\Delta t,t)+\sum_{u\ne s}(\lambda_up_{us}\Delta t) n_u(x-v_u\hat\Delta t-v_s(\Delta t-\hat\Delta t),t)+o(\Delta t),$$
which can be further simplified as
$$\begin{aligned}
    n_s(x,t)+(\Delta t)\partial_t n_s(x,t) +o(\Delta t)=&\ (1-\lambda_s\Delta t)\left(n_s(x,t)-(v_s\Delta t )\partial_x n_s(x,t) + o(\Delta t)\right) +
    \\
    &\sum_{u\ne s}(\lambda_up_{us}\Delta t) (n_u(x,t)-(v_s\Delta t+(v_u-v_s)\hat\Delta t)\partial_x n_u(x,t)+o(\Delta t))
    \\
    &+o(\Delta t).
\end{aligned}$$
Since $\hat\Delta t< \Delta t$, we have $\Delta t\cdot \hat \Delta t = o(\Delta t)$, and using $(\Delta t)^2 = o(\Delta t)$, we write
$$\begin{aligned}
    n_s(x,t)+(\Delta t)\partial_t n_s(x,t) +o(\Delta t)=&\ (1-\lambda_s\Delta t)n_s(x,t)-(v_s\Delta t )\partial_x n_s(x,t) +o(\Delta t) + o(\Delta t)
    \\
    &+\sum_{u\ne s}(\lambda_up_{us}\Delta t) n_u(x,t) +o(\Delta t) +o(\Delta t)
    \\
    &+o(\Delta t).
\end{aligned}$$
By grouping the terms $o(\Delta t)$ and rearranging, we obtain
$$(\Delta t)\partial_t n_s =  -(v_s\Delta t )\partial_x n_s(x,t)+(-\lambda_s\Delta t)n_s(x,t)+\sum_{u\ne s}(\lambda_up_{us}\Delta t) (n_u(x,t))+o(\Delta t).$$
To conclude, we divide by $\Delta t$ and let $\Delta t\to0$ to obtain the three-state PDE model
\begin{equation}\label{SI:Eq:n-state_RA}
    \partial_t n_s(x,t)+v_s\partial_x n_s(x,t)= \sum_{u=1}^3 q_{us}n_u(x,t).
\end{equation}
In this work, we assume that we can measure only the total particle density
$$N(x,t)=\sum_{s=1}^3 n_s(x,t).$$ 
For brevity, we denote $n_s=n_s(x,t)$ and $N=N(x,t)$.

\section{Structural identifiability measuring single-particle trajectories if two states share the same velocity}\label{SI:Sec:SI_IBM_v2=v3}

In this section, we analyse the structural identifiability properties of a three-state model in which two states share the same velocity and, without loss of generality, we assume that $v_1\ne v_2=v_3=:v$. The velocities $v_1$ and $v$ can be identified by finding all values taken by $x'(t)$. The lengths of all time intervals spent with velocity $\nu$, denoted $\tau_{\nu}^1,\tau_{\nu}^2,\ldots$,  can be used to write the cumulative waiting-time distribution of the time spent in all segments with constant velocity $\nu$ before switching
$$\mathbb{P}(t_{\nu}\le t)=\lim_{K\to\infty}\frac{1}{K}\sum_{k=1}^K \mathds{1} (\tau_{\nu}^k\le t),$$
and use it to identify $\lambda_1$ as $\mathbb{P}(t_{v_1}> t)=\exp{(-\lambda_1 t)}$.

In contrast, an observed interval with velocity $v$ may contain multiple
unobserved transitions between states $s=2$ and $s=3$ as the individual states cannot be distinguished from the observed trajectory. We therefore study the duration of an observed interval with velocity $v$ before the process returns to velocity $v_1$. 

Let $T_{23}$ denote the duration of an observed interval with velocity $v$ after switching from velocity $v_1$. Since both states $s=2$ and $s=3$ share the same velocity, we cannot distinguish their initial state. Therefore, to find the distribution of $T_{23}$ we need to consider the two conditional survival probabilities
$$R_2(t)
    =
    \mathbb{P}(T_{23}>t
    \mid \text{the interval begins in state }s=2),$$
and
$$R_3(t)=\mathbb{P}(T_{23}>t
    \mid \text{the interval begins in state }s=3).$$
Then, the observed survival probability function of a velocity-$v$ interval is
\begin{equation}\label{Eq:tau_23_definition}
    \tau_{23}(t)
    :=
    \mathbb{P}(T_{23}>t)=p_{12}R_{2}(t)+p_{13}R_{3}(t),
\end{equation}
since, after leaving state $s=1$, the process enters state $s=2$ with probability $p_{12}$ and state $s=3$ with probability $p_{13}$. 

Now, we write a system of ODEs that allows us to find an expression for $R_2$ and $R_3$. Conditioning on the behaviour of the process over an arbitrarily small time interval $[0,\Delta t]$, we can write $R_2(t+\Delta t)$ as the survival probability of staying in state $s=2$, plus the probability of switching to state $s=3$, and some $o(\Delta t)$ terms. Therefore, we write
$$
    R_2(t+\Delta t)
    =
    (1-\lambda_2\Delta t)R_2(t)
    +
    \lambda_2p_{23}\Delta t\,R_3(t)
    +
    o(\Delta t),
$$
and, analogously,
$$
    R_3(t+\Delta t)
    =
    (1-\lambda_3\Delta t)R_3(t)
    +
    \lambda_3p_{32}\Delta t\,R_2(t)
    +
    o(\Delta t).
$$
Rearranging, dividing by $\Delta t$, and taking the limit $\Delta t\to0$, we obtain the equations
\begin{equation}
    R_2'(t)
    =
    -\lambda_2R_2(t)
    +
    \lambda_2p_{23}R_3(t),
    \label{Eq:R2_backward}
\end{equation}
and
\begin{equation}
    R_3'(t)
    =
    \lambda_3p_{32}R_2(t)
    -
    \lambda_3R_3(t),
    \label{Eq:R3_backward}
\end{equation}
with $R_2(0)=R_3(0)=1$. This system of ODEs can be written as
\begin{equation}\label{Eq:SYS_ODE_R}
\mathbf{R}'(t)=\mathbf{T}\mathbf{R}(t),
    \qquad
    \mathbf{R}(0)=\mathbf{1},
\end{equation}
where
$$\mathbf{R}(t)
    =
    \begin{bmatrix}
        R_2(t)\\
        R_3(t)
    \end{bmatrix},
    \qquad
    \mathbf{T}
    =
    \begin{bmatrix}
        -\lambda_2 & \lambda_2p_{23}\\
        \lambda_3p_{32} & -\lambda_3
    \end{bmatrix}.$$
    
The characteristic polynomial of $\mathbf{T}$ is
$$
    \chi_\mathbf{T}(r)= r^2+k_1r+k_2,
$$
where $k_1=\lambda_2+\lambda_3,$ and $k_2=\lambda_2\lambda_3
    \left(
        p_{21}+p_{31}-p_{21}p_{31}
    \right).$ Therefore, by the Cayley-Hamilton theorem~\cite{horn2012matrix}, $\mathbf{T}$ satisfies
$$
\mathbf{T}^2+k_1\mathbf{T}+k_2\mathbf{I}=0,
$$
where $\mathbf{I}$ is a $2\times 2$ identity matrix. Since $\mathbf{R}''(t)=\mathbf{T}^2\mathbf{R}(t)$, Equation~\eqref{Eq:tau_23_definition} implies
\begin{equation}
    \tau_{23}''(t)
    +
    k_1\tau_{23}'(t)
    +
    k_2\tau_{23}(t)
    =
    0,
    \label{Eq:tau23_scalar_ODE}
\end{equation}
with initial conditions
$$
    \tau_{23}(0)=1, \qquad
    \tau_{23}'(0)
    =-k_3,
$$
where $ k_3= \lambda_2p_{12}p_{21}
+ \lambda_3(1-p_{12})p_{31}.$

Thus, the observed waiting-time distribution depends only on \begin{equation}\label{Eq:k1-k3_v2=v3}
k_1=\lambda_2+\lambda_3, \ \ k_2=\lambda_2 \lambda_3 (p_{21}+p_{31}-p_{21}p_{31}), \ \ k_3=\lambda_2 p_{12} p_{21}+\lambda_3 p_{31}(1 - p_{12}).
\end{equation}
Since the initial condition $\tau_{23}'(0)=-k_3$ is observed, the parameter $k_3$ is structurally identifiable. Moreover, the waiting-time distribution is observed; therefore, Equation~\eqref{Eq:tau23_scalar_ODE} is an input-output equation, which gives that $k_1,k_2$ are also structurally identifiable. However, the three identifiable parameter combinations $k_1,k_2,k_3$ are not sufficient to identify the five parameters $\lambda_2,\lambda_3,p_{12},p_{21},p_{31}$. Therefore, the three-state model is not structurally identifiable when two states share the same velocity.

We now consider the initial condition. As in Section~3 of the main text, we observe that the function $f(x)$ is structurally identifiable by measuring the initial locations of an infinite number of particles. Moreover, $a_1$ can be identified as
$$a_1 = \lim_{M\to\infty} \frac{1}{M} \sum_{m=1}^M \mathds{1} (x_m'(0)=v_1),$$
where we exclude the particles for which $x_m$ is not differentiable at zero, and
$$a_2+a_3
    =
    \lim_{M\to\infty}
    \frac{1}{M}
    \sum_{m=1}^{M}
    \mathds{1} \left(x_m'(0)=v\right)=1-a_1,$$
is also identifiable. However, the initial weights $a_2$ and $a_3$ cannot be identified separately from the initial velocities. The only exception occurs when $a_1=1$, in which case we can also identify $a_2=a_3=0$, thus, now we focus on the case $a_1<1$.

We study whether there is additional identifiability information that can be obtained from the type of initial condition considered. Therefore, we study again the waiting-time distribution of the intervals with velocity $v$. Using the notation as above, we write the initial survival probability $\tau_{23}^{(0)}(t)$ using the conditional survival probabilities $R_2(t)$ and $R_3(t)$ as
\begin{equation}\label{Eq:tau230}
    \tau_{23}^{(0)}(t)
    =
    \frac{a_2}{1-a_1}R_2(t)
    +
    \frac{a_3}{1-a_1}R_3(t),
\end{equation}
since $a_2/(1-a_1)$ is the probability of a particle being initially in state $s=2$, when its velocity is $v$, and similarly for state $s=3$. Since $R_2$ and $R_3$ satisfy the system in Equation~\eqref{Eq:SYS_ODE_R},
$\tau_{23}^{(0)}$ satisfies
\begin{equation}
    \frac{\mathrm{d}^2\tau_{23}^{(0)}}{\mathrm{d}t^2}
    +
    k_1\frac{\mathrm{d}\tau_{23}^{(0)}}{\mathrm{d}t}
    +
    k_2\tau_{23}^{(0)}
    =
    0,
    \label{Eq:tau23_initial_ODE}
\end{equation}
with $\tau_{23}^{(0)}(0)=1$ and, from the derivative of Equation~\eqref{Eq:tau230} at $t=0$,
$$
    \left.
    \frac{\mathrm{d}\tau_{23}^{(0)}}{\mathrm{d}t}
    \right|_{t=0}
    =
    -
    \frac{
        a_2\lambda_2p_{21}
        +
        a_3\lambda_3p_{31}
    }{
        1-a_1
    }.
$$
Thus, since $a_1$ is identifiable and $a_3=1-a_1-a_2$, the derivative of the initial waiting-time distribution $\tau_{23}^{(0)}$ at time $t=0$ identifies the additional parameter combination
\begin{equation}\label{Eq:k4}
    k_4=(1-a_1)\lambda_3p_{31}
    +a_2\left(
        \lambda_2p_{21}-\lambda_3p_{31}
    \right).
\end{equation}
Therefore, the function $f(x)$ and the parameters $a_1$ and $k_4$ are identifiable. To conclude, the identifiable quantities are
$v_1,v,\lambda_1,a_1,k_1,k_2,k_3,k_4$, but the parameters
$\lambda_2,\lambda_3,p_{12},p_{21}$, $p_{31},a_2$ are not structurally identifiable.

\section{Structural identifiability measuring the total particle density when two states share the same velocity}\label{SI:Sec:SI_PDE_v2=v3}

In this section, we study the structural identifiability of the model from total particle density data, when two states share the same velocity, and without loss of generality, we assume $v_1\ne v_2=v_3=:v$. A two-state model in which both states share the same velocity is trivial; therefore, this is the simplest non-trivial case in which distinct internal states cannot be distinguished from the observed velocity. Since $v_1\ne v$, the differential algebra steps in Section~4.1 to obtain the input-output equation are still valid, and the coefficients obtained are
\begin{equation}\label{SI:Eq:c1-c8_v2=v3}
\begin{aligned}
    &\hat c_1=v_1 + 2v,\qquad \hat c_2=2v_1 v + v^2,\quad \hat c_3=v_1 v^2,
    \\
    &\hat c_4=\lambda_1 +(\lambda_2 + \lambda_3),\qquad \hat c_5= 2\lambda_1 v + (\lambda_2 +\lambda_3) (v_1 + v),
    \\
    &\hat c_6= \lambda_1 v^2 + (\lambda_2+\lambda_3) v_1 v,
    \\
    & \hat c_7=\lambda_2 \lambda_3 (1-(1-p_{21})(1-p_{31})) + \lambda_1 [\lambda_3 (1 - (1 - p_{12}) p_{31}) + \lambda_2 (1 - p_{12} p_{21})], 
    \\
    & \hat c_8=\lambda_2 \lambda_3 (1-(1-p_{21})(1-p_{31})) v_1 \\
    &\qquad + \lambda_1 [\lambda_3 (1 - (1 - p_{12}) p_{31})
     + \lambda_2 (1 - p_{12} p_{21})] v.
\end{aligned}
\end{equation}
The identifiability up to state relabelling of the velocities still holds. The coefficients $\hat c_4,\hat c_5$ give identifiability of the rate $\lambda_1$ and of the sum $\lambda_2+\lambda_3$, as $v\ne v_1$, and the coefficient $\hat c_6$ is redundant. Then, the remaining coefficients $\hat c_7,\hat c_8$, simplified for known quantities, together with $\lambda_2+\lambda_3$, give the identifiable parameter combinations 
\begin{equation}\label{Eq:k1-k3_v2=v3}
k_1=\lambda_2+\lambda_3, \ \ k_2=\lambda_2 \lambda_3 (p_{21}+p_{31}-p_{21}p_{31}), \ \ k_3=\lambda_2 p_{12} p_{21}+\lambda_3 p_{31}(1 - p_{12}).
\end{equation}
These three parameter combinations are not sufficient to identify the five remaining parameters $\lambda_2,\lambda_3,p_{12},p_{21},p_{31}$.

\section{Examples of equivalent model parametrisations and the impact of the initial condition with velocities all distinct}

In this section, we study how the initial conditions of the type $n_s(x,0)=a_s f(x)$ impact the structural identifiability of the three-state PDE model, by comparing the total density evolutions obtained for models with the same coefficients $c_1,c_2,\ldots,c_9$. Now, we showcase examples of models with the same input-output coefficients but different probability parameters.

In the examples provided, the numerical simulations of the PDE model solutions are obtained with a first-order upwind scheme for the spatial discretisation, setting the spatial grid to $\text{d}x=0.01$. Then, the time integration is obtained with the automatic method switching discretisation LSODA~\citep{petzold1983automatic}, with output time discretisation with maximal step $\text{d}t=0.00045$, which ensures that the Courant–Friedrichs–Lewy stability condition is satisfied~\citep{courant1967partial}. For tractability of the numerical solutions, we use periodic boundary conditions in the domain $[-L,L]$ for some sufficiently large $L>t\max_s{|v_s|}$, such that $N(x,t)$ is approximately zero close to the boundaries for the time interval considered. Indeed, we define the initial condition $f(x)$ as the Gaussian distribution in the interval $[-L,L]$, with zero mean and variance $\varsigma^2/2$,
\begin{equation}\label{SI:Eq:f(x)}
    f(x)=\frac{1}{\sqrt{\pi\varsigma^2}}\exp{\left(-\frac{x^2}{\varsigma^2}\right)},
\end{equation}
with $\varsigma=5$, and $L=40$, and we extend it by periodicity to $\mathbb{R}$.

\subsection{The coefficients \texorpdfstring{$c_1,\ldots,c_9$}{c1,...,c9} do not capture all information from the separable initial condition}\label{App:three-state_PDE_c1-c9}

\begin{figure*}[!ht]
    \centering
    \includegraphics[width=0.7\textwidth]{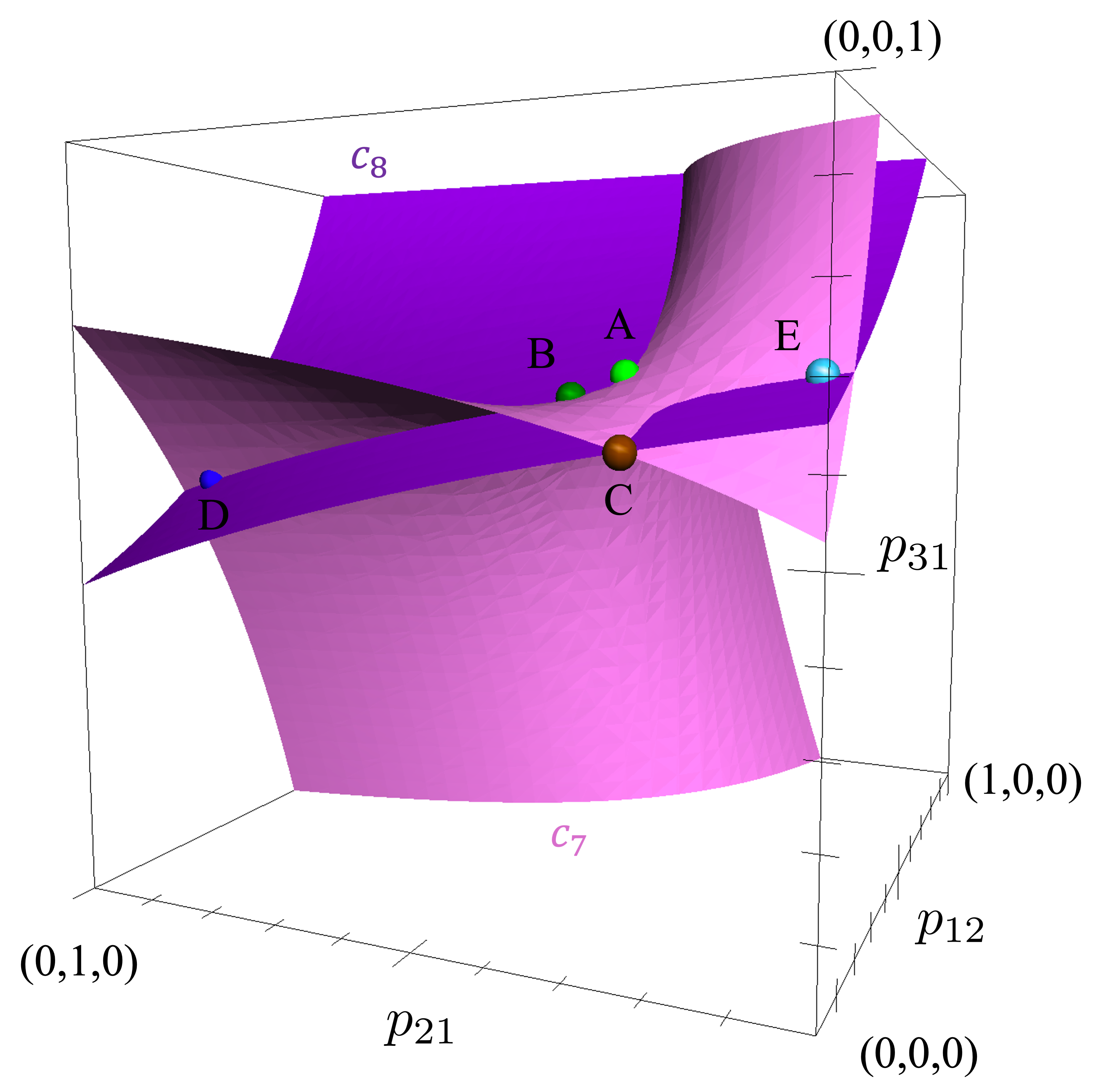}
    \caption[Example surfaces obtained fixing the coefficients $c_7$ and $c_8$]{\textbf{Example surfaces obtained fixing the coefficients $c_7$ and $c_8$.} The surfaces shown are obtained fixing the coefficients $c_7=1441/2000$ (pink) and $c_8=39/100$ (violet), obtained from $\boldsymbol{\theta}_A$, assuming identified parameters $\boldsymbol{\theta}=\{v_1=20, v_2=-15, v_3=0, \lambda_1=1, \lambda_2=1/2, \lambda_3=3/10\}$ and varying $(p_{12},p_{21},p_{31})\in[0,1]\times[0,1]\times[0,1]$. The graph also shows the points related to the parameter sets $\boldsymbol{\theta}_A$ (light green), $\boldsymbol{\theta}_B$ (dark green), $\boldsymbol{\theta}_C$ (brown), $\boldsymbol{\theta}_D$ (blue) and $\boldsymbol{\theta}_E$ (light blue). This image was created using Apple Grapher (macOS).}
    \label{SI:Figure_S1}
\end{figure*}

Now, we show that the input-output coefficients $c_1,\ldots,c_9$ are not sufficient to fully establish the structural identifiability properties of the three-state models. In particular, we show examples with identical and different total particle density evolution, but different model parameters (Figure~\ref{SI:Figure_S1}). First, without loss of generality, we define a reference parameter set $\boldsymbol{\theta}_A$. Then, we choose parameter sets $\boldsymbol{\theta}_B$ and $\boldsymbol{\theta}_C$ with same coefficients $c_1,\ldots,c_9$ as $\boldsymbol{\theta}_A$. In particular, $\boldsymbol{\theta}_B$ is the only parametrisation with the same stationary distribution as $\boldsymbol{\theta}_A$, while $\boldsymbol{\theta}_C$ is chosen to have $p_{12}=0$. These examples are also used to analyse the impact of the initial condition on the total density evolution and, therefore, on their structural identifiability properties.

The parameter sets considered have the common parameters
\begin{equation}\label{Eq:theta}
    \boldsymbol{\theta}=\{v_1=20, v_2=-15, v_3=0, \lambda_1=1, \lambda_2=1/2, \lambda_3=3/10\},
\end{equation}
fixed by $c_1,\ldots,c_6$. Without loss of generality, we choose
$$\boldsymbol{\theta}_{A}= \boldsymbol{\theta}\cup \{p_{12}=1/5, p_{21}=3/10, p_{31}=7/10\}.$$
The parameters of the three-state models with the same coefficients $c_1,\ldots,c_8$ as $\boldsymbol{\theta}_{A}$ must lie in the intersection of the surfaces obtained by fixing the coefficients $c_7$ and $c_8$ (Figure~\ref{SI:Figure_S1}; pink and violet surfaces), given by
\begin{equation}\label{SI:Eq:example_2_coefficients}
\begin{dcases}
    \frac{1441}{2000}=\frac{3}{20}(1-(1-p_{21})(1-p_{31})) + \frac{3}{10}(1 - (1 - p_{12}) p_{31}) + \frac{1}{2}(1 - p_{12} p_{21}), 
    \\
    \frac{39}{100}=3(1-(1-p_{21})(1-p_{31})) -\frac{9}{2} (1 - (1 - p_{12}) p_{31}).
\end{dcases}
\end{equation}
We note that there are an infinite number of other sets of probabilities that give the same coefficients in the input-output equation.

Then, the parameter set $\boldsymbol{\theta}_B$ is chosen to also have the same stationary distribution as $\boldsymbol{\theta}_A$, and $\boldsymbol{\theta}_C$ is chosen to have $p_{12}=0$. We choose the parameter sets
\begin{equation*}
\begin{aligned}
    \boldsymbol{\theta}_{B} &= \boldsymbol{\theta}\cup \{p_{12}=66/395, p_{21}=79/220, p_{31}=158/235\},
    \\
    \boldsymbol{\theta}_{C} &= \boldsymbol{\theta}\cup \{p_{12}=0, p_{21}=61/268, p_{31}=108/175\},
\end{aligned}
\end{equation*}
and the stationary distribution weights are
\begin{equation*}
    \boldsymbol{\pi}_{A}=\boldsymbol{\pi}_{B}=[237/1441, 24/131, 940/1441]^T,
\end{equation*}
and
\begin{equation*}
    \boldsymbol{\pi}_{C}=[1479/10087, 1608/10087, 1000/1441]^T.
\end{equation*}
Now, we study the impact of the initial conditions by comparing numerical solutions for the parameter sets $\boldsymbol{\theta}_A$, $\boldsymbol{\theta}_B$ and $\boldsymbol{\theta}_C$, obtained fixing the coefficients $c_1,\ldots,c_8$, for some initial condition weights that fix the coefficient $c_9$.

\begin{figure*}[!ht]
    \centering
    \includegraphics[width=\textwidth]{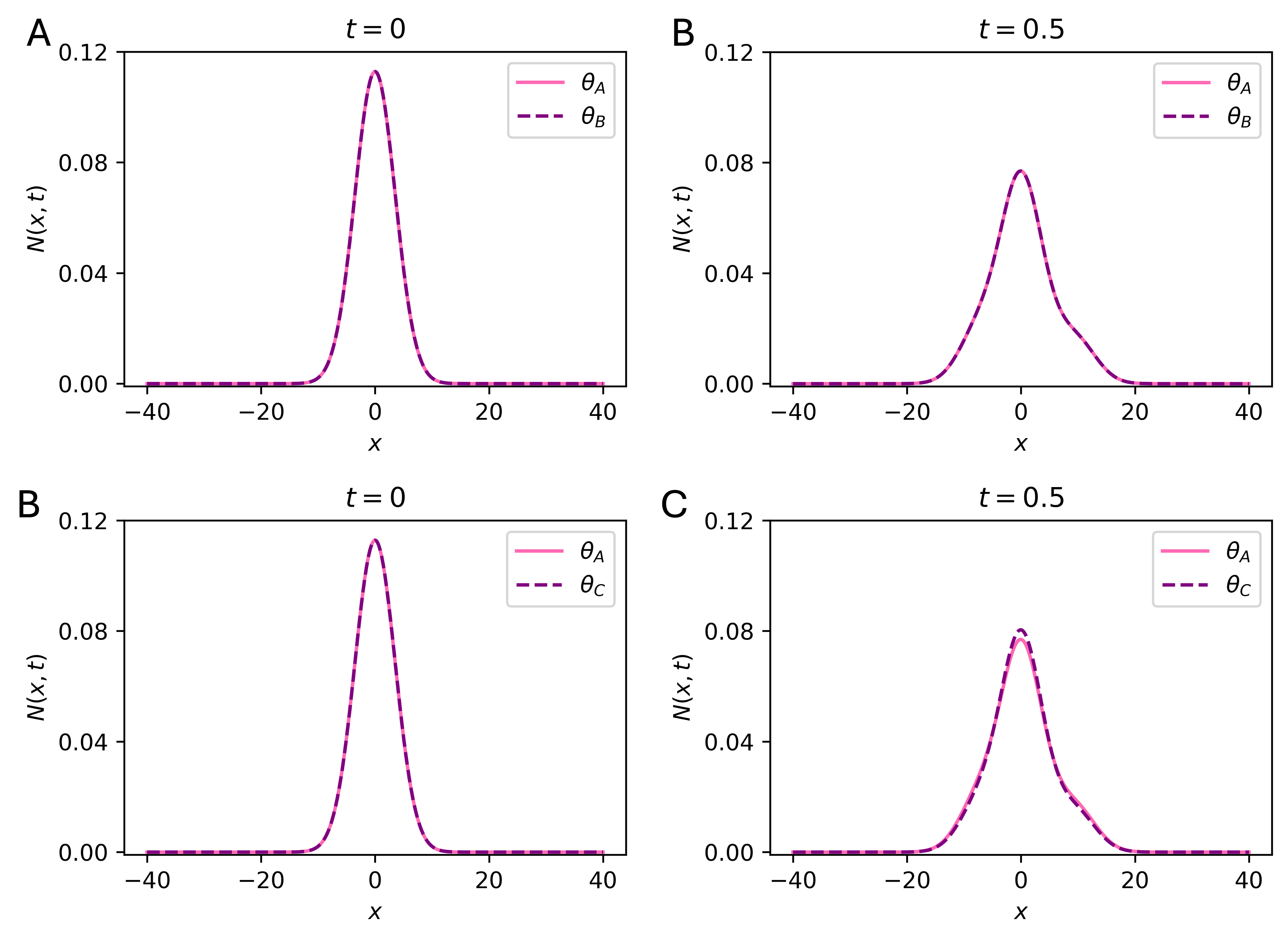}
    \caption[Numerical solution for $\boldsymbol{\theta}_A$, $\boldsymbol{\theta}_B$ and $\boldsymbol{\theta}_C$]{\textbf{Numerical solution for $\boldsymbol{\theta}_A$, $\boldsymbol{\theta}_B$ and $\boldsymbol{\theta}_C$ with initial conditions $\boldsymbol{a}=\boldsymbol{\pi}$, at times $t=0$ and $t=0.5$.} The numerical solution for $\boldsymbol{\theta}_A$ is compared with the one for $\boldsymbol{\theta}_B$ and $\boldsymbol{\theta}_C$, with $\boldsymbol{a}=\boldsymbol{\pi}$, at times $t=0$ and $t=0.5$.}
    \label{Fig_2_AvsBC}
\end{figure*}

Firstly, we fix the initial state distribution $\boldsymbol{a}=[a_1,a_2,1-a_1-a_2]^T$ to be equal to the stationary distribution $\boldsymbol{\pi}$ (see Section~\ref{Sec:S1}). From Figure~\ref{Fig_2_AvsBC}, we observe that the total particle density evolution is identical for $\boldsymbol{\theta}_A$ and $\boldsymbol{\theta}_B$ (panels \textbf{A}-\textbf{B}), while it is different between $\boldsymbol{\theta}_A$ and $\boldsymbol{\theta}_C$ (panels \textbf{C}-\textbf{D}), despite having equal coefficients $c_1,\ldots,c_9$. 

\begin{figure*}[!ht]
    \centering
    \includegraphics[width=\textwidth]{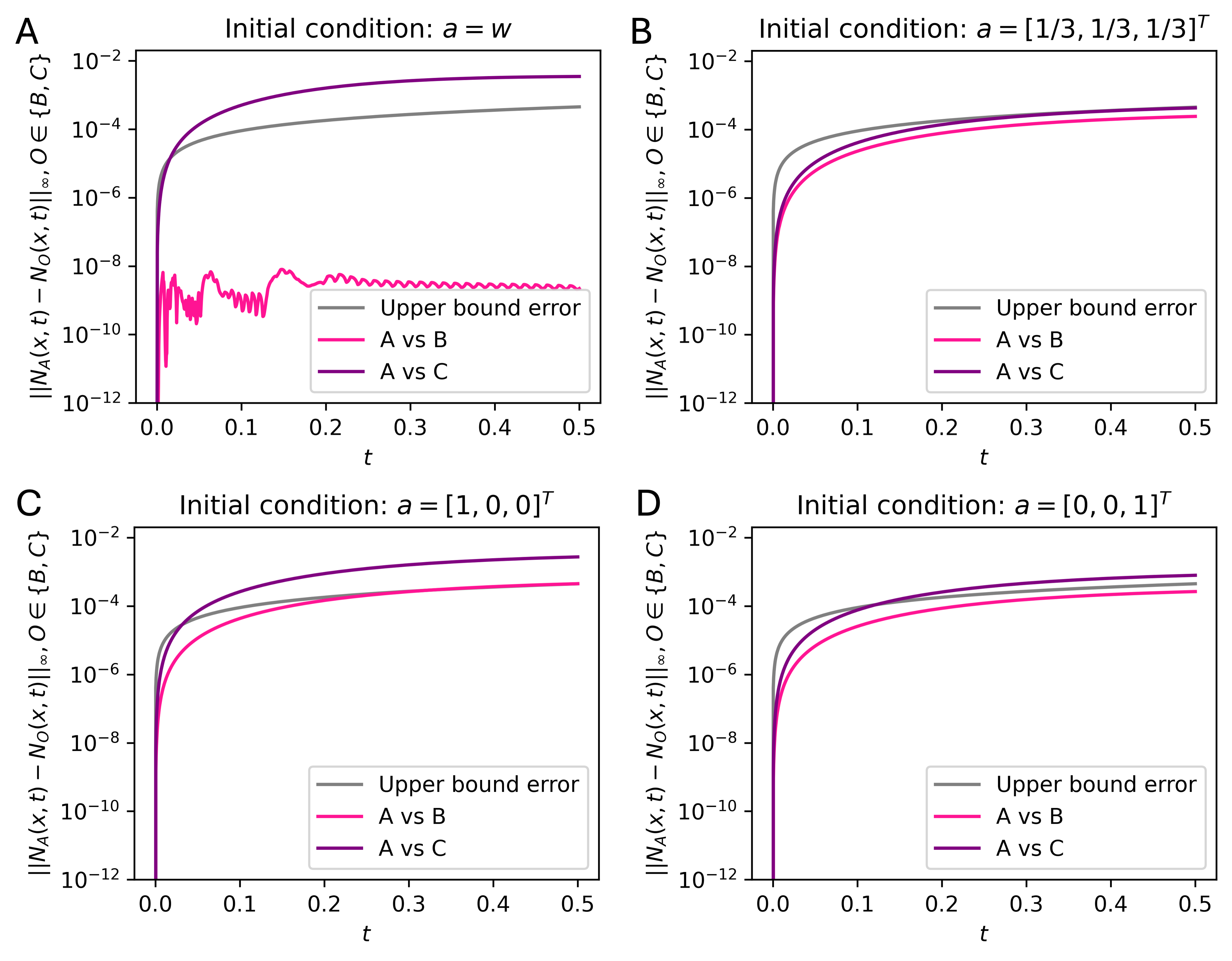}
    \caption[L-infinity norm of the difference of $N(x,t)$ for $\boldsymbol{\theta}_A$ versus $\boldsymbol{\theta}_B$ and $\boldsymbol{\theta}_C$]{\textbf{L-infinity norm of the difference of $N(x,t)$ for $\boldsymbol{\theta}_A$ versus $\boldsymbol{\theta}_B$ and $\boldsymbol{\theta}_C$, as a function of time $t\in[0,T]$, where $T=0.5$.} L-infinity norm of the difference of $N(x,t)$ between $\boldsymbol{\theta}_A$ and $\boldsymbol{\theta}_B$ and $\boldsymbol{\theta}_C$ with distinct initial condition weights: \textbf{A} $\boldsymbol{a}=\boldsymbol{\pi}$, \textbf{B} $\boldsymbol{a}=[1/3,1/3,1/3]^T$, \textbf{C} $\boldsymbol{a}=[1,0,0]^T$, and \textbf{D} $\boldsymbol{a}=[0,0,1]^T$. The upper bound of the error is computed as infinity norm of the difference between the numerical simulation of the PDE $\partial_tN = \max_s |v_s|\cdot \partial_x N$ using the upwind scheme with LSODA and its analytical solution.}
    \label{SI:Fig_S3}
\end{figure*}

Further demonstration of the difference between the total density evolution of these two models can be seen by comparing their error to an upper bound of the numerical error for the upwind scheme used. This is obtained as the error difference between the exact and numerical solutions for a one-state model with velocity $v_1=20$. Panel \textbf{A} in Figure~\ref{SI:Fig_S3} shows errors of an order approximately between $10^{-10}$ and $10^{-7}$ for $\boldsymbol{\theta}_A$ versus $\boldsymbol{\theta}_B$, while errors of an order between $10^{-4}$ and $10^{-2}$ for $\boldsymbol{\theta}_A$ versus $\boldsymbol{\theta}_C$, which are close to or higher than the upper bound of the numerical error obtained ($10^{-4}$). The different total density evolution between $\boldsymbol{\theta}_A$ and $\boldsymbol{\theta}_C$ suggests that the initial conditions carry information on the model parameters that is not captured by the coefficients $c_1,\ldots,c_9$.

To further analyse the behaviour of the model solutions, we repeat the same analysis with separable initial condition, with unknown weights set to $\boldsymbol{a}=[1/3,1/3,1/3]^T$, $\boldsymbol{a}=[1,0,0]^T$, and $\boldsymbol{a}=[0,0,1]^T$. These choices of initial conditions correspond to assuming that particles at time $t=0$ are equally distributed across states, or are all in state $s=1$, with velocity $v_1=20$, or are all in state $s=3$, with velocity $v_3=0$, respectively. Figure~\ref{SI:Fig_S3}\textbf{B}-\textbf{D} compares the evolution of the error difference of total density $N(x,t)$, for $\boldsymbol{\theta}_A$ to $\boldsymbol{\theta}_B$ and $\boldsymbol{\theta}_C$, with initial weights $\boldsymbol{a}=[1/3,1/3,1/3]^T$, $\boldsymbol{a}=[1,0,0]^T$, and $\boldsymbol{a}=[0,0,1]^T$. In each of these cases, we observe that, despite choosing the initial condition weights to be equal across models, the error between the total density evolutions is still large (around $10^{-4}$ or higher). 

\begin{figure*}[!ht]
    \centering
    \includegraphics[width=\textwidth]{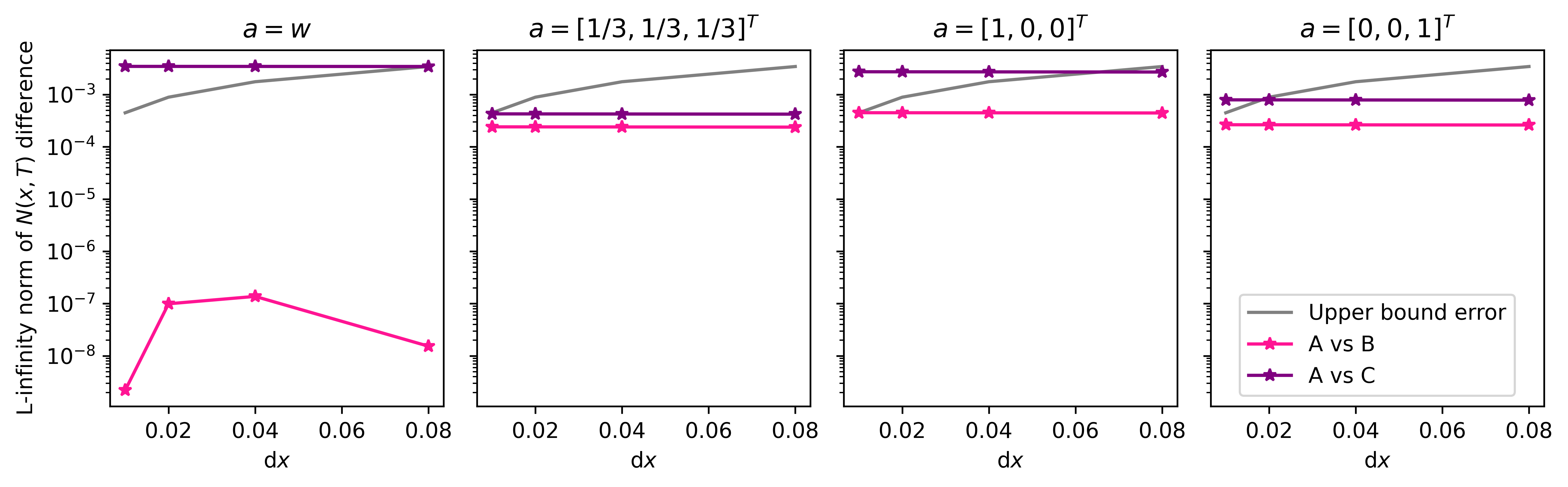}
    \caption[L-infinity norm of the difference of $N(x,T)$ between $\boldsymbol{\theta}_A$, $\boldsymbol{\theta}_B$ and $\boldsymbol{\theta}_C$]{\textbf{L-infinity norm of the difference of $N(x,T)$ between $\boldsymbol{\theta}_A$ and $\boldsymbol{\theta}_B$ and $\boldsymbol{\theta}_C$ at $T=0.5$.} The L-infinity norms of
    $N(x,T|\boldsymbol{\theta}_{A})-N(x,T|\boldsymbol{\theta}_{B})$ (pink), 
    $N(x,T|\boldsymbol{\theta}_{A})-N(x,T|\boldsymbol{\theta}_{C})$ (purple) are shown. The $x$ grid increment $\text{d}x$ is varied in the set $\{0.01, 0.02, 0.04, 0.08\}$, and the $\text{d}t$ is also varied to guarantee numerical stability of the PDE solver. The grey line represents an upper bound for the numerical error as the grid is varied.}
    \label{SI:Fig_S4}
\end{figure*}

Now, we study how the error varies with the grid resolution, varying $\text{d} x$, for four types of initial conditions, with weights $\boldsymbol{a}=\boldsymbol{\pi}$, $\boldsymbol{a}=[1/3,1/3,1/3]^T$, $\boldsymbol{a}=[1,0,0]^T$ and $\boldsymbol{a}=[0,0,1]^T$. In Figure~\ref{SI:Fig_S4} we show the L-infinity norm of the difference of the total density evolution between $\boldsymbol{\theta}_A$, $\boldsymbol{\theta}_B$ and $\boldsymbol{\theta}_C$ at time $T=0.5$, and we interpret the norms close to or above the upper bound error to be significant, and the significantly lower values to be numerical artefacts. We observe that only $\boldsymbol{\theta}_A$ and $\boldsymbol{\theta}_B$ with the same initial condition $\boldsymbol{a}=\boldsymbol{\pi}=\boldsymbol{\pi}_A=\boldsymbol{\pi}_B$ produce the same total particle density evolution.

Overall, these results demonstrate that fixing the coefficients $c_1,\ldots,c_9$ is not sufficient to determine the total particle density evolution. Therefore, the additional information arising from the specified initial condition remains to be incorporated into the structural identifiability analysis.

\subsection{Structurally equivalent probability parameters depend on the initial condition parameters}

We now consider how the structurally equivalent parameter sets depend on the specified initial condition. This highlights the connection between experimental design and structural identifiability and suggests that, when experimentally feasible, varying the initial condition may provide additional information for distinguishing between otherwise equivalent parameter sets. The parameter sets $\boldsymbol{\theta}_A$ and $\boldsymbol{\theta}_B$, with initial condition weights $\boldsymbol{a}=\boldsymbol{\pi}$ and different probability parameters $p_{12},p_{21},p_{31}$, have the same coefficients $c_1,\ldots,c_{11}$ and give the same total particle density evolution (Figure~\ref{Fig_2_AvsBC}\textbf{B}). 

\begin{figure*}[!ht]
    \centering
    \includegraphics[width=\textwidth]{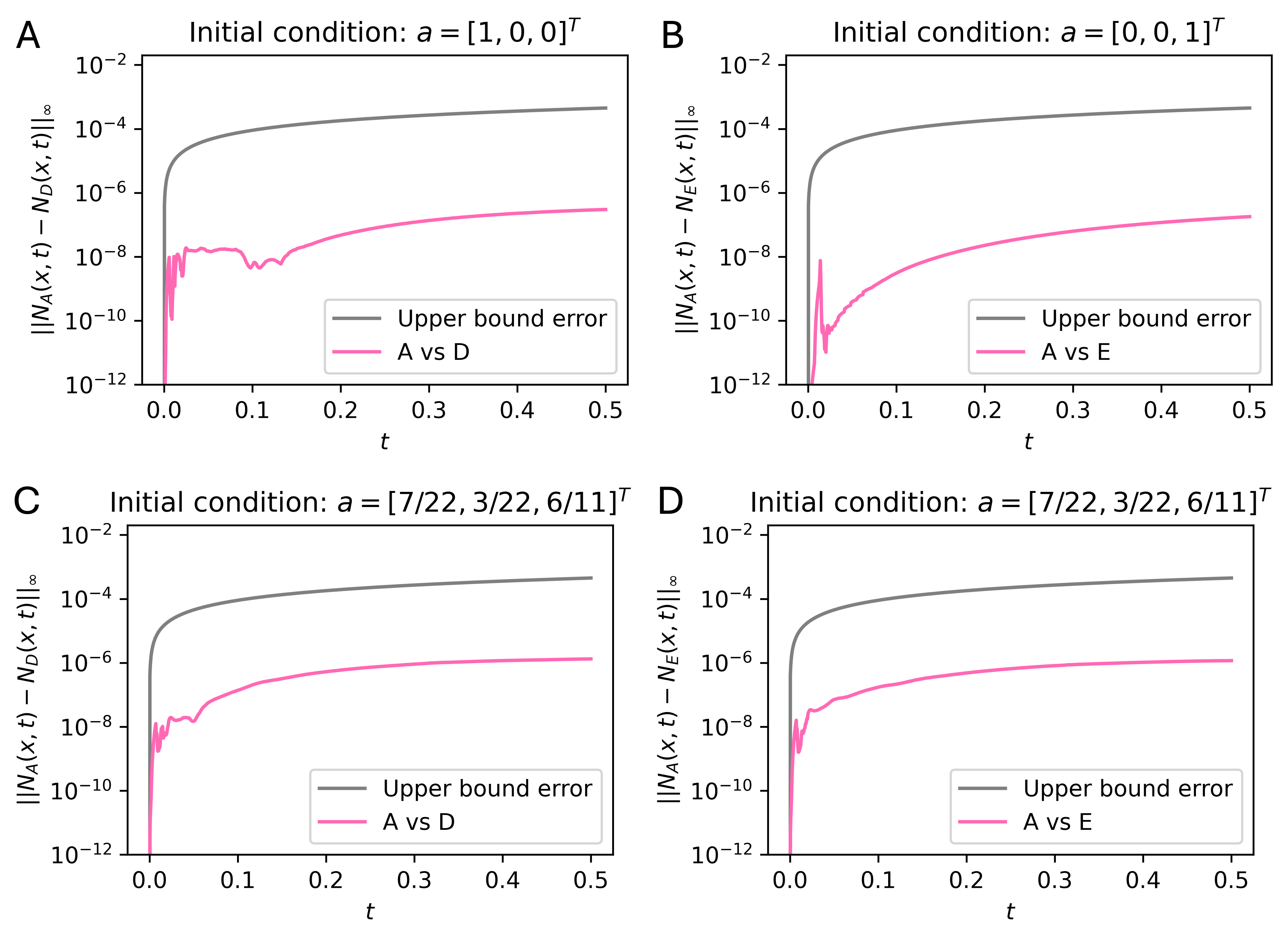}
    \caption[L-infinity norm of the difference of $N(x,t)$ between $\boldsymbol{\theta}_A$ and $\boldsymbol{\theta}_D$ and $\boldsymbol{\theta}_E$]{\textbf{L-infinity norm of the difference of $N(x,t)$ between $\boldsymbol{\theta}_A$, $\boldsymbol{\theta}_D$ and $\boldsymbol{\theta}_E$ as a function of time $t\in[0,0.5]$}. \textbf{A} L-infinity norm of $N(x,t|\boldsymbol{\theta}_A)-N(x,t|\boldsymbol{\theta}_D)$ with initial condition weights $\boldsymbol{a}=[1,0,0]^T$. \textbf{B} L-infinity norm of $N(x,t|\boldsymbol{\theta}_A)-N(x,t|\boldsymbol{\theta}_E)$ with initial condition weights $\boldsymbol{a}=[0,0,1]^T$. \textbf{C} L-infinity norm of $N(x,t|\boldsymbol{\theta}_A)-N(x,t|\boldsymbol{\theta}_D)$ with initial condition weights $\boldsymbol{a}=[7/22, 3/22, 6/11]^T$. \textbf{D} L-infinity norm of $N(x,t|\boldsymbol{\theta}_A)-N(x,t|\boldsymbol{\theta}_E)$ with initial condition weights $\boldsymbol{a}=[7/22, 3/22, 6/11]^T$.}
    \label{SI:Figure_S5}
\end{figure*}

The only physically admissible parameter set structurally equivalent to $\boldsymbol{\theta}_A$ with initial condition $\boldsymbol{a}_A=\boldsymbol{\pi}_A$ is $\boldsymbol{\theta}_B$ with initial condition $\boldsymbol{a}_A=\boldsymbol{\pi}_B$, since $\boldsymbol{\pi}_A=\boldsymbol{\pi}_B$. For $\boldsymbol{\theta}_A$ with initial condition $\boldsymbol{a}=[1/3,1/3,1/3]^T$ there is no other physically admissible structurally equivalent parameter set, making this parameter set globally structurally identifiable. For $\boldsymbol{\theta}_A$ with initial condition $\boldsymbol{a}=[1,0,0]^T$ there is one structurally equivalent parameter set (Figure~\ref{SI:Figure_S5}\textbf{A})
$$\boldsymbol{\theta}_{D}=\boldsymbol{\theta}\cup \{p_{12}=1/5, p_{21}=47/50, p_{31}=23/42\},$$
with $\boldsymbol{\theta}$ as specified in Equation~\eqref{Eq:theta}. Similarly, when $\boldsymbol{\theta}_A$ has initial condition weights $\boldsymbol{a}=[0,0,1]^T$, there is one structurally equivalent parameter set (Figure~\ref{SI:Figure_S5}\textbf{B})
$$\boldsymbol{\theta}_{E}=\boldsymbol{\theta}\cup \{p_{12}=87/700, p_{21}=7/200, p_{31}=7/10\}.$$

We also note that $\boldsymbol{\theta}_A$ with initial condition $\boldsymbol{a}=[7/22, 3/22, 6/11]^T$ is equivalent to both $\boldsymbol{\theta}_D$, and $\boldsymbol{\theta}_E$ with the same initial condition (Figure~\ref{SI:Figure_S5}\textbf{C}-\textbf{D}). The equivalent parameter sets $\boldsymbol{\theta}_A, \boldsymbol{\theta}_D,\boldsymbol{\theta}_E$ with initial condition $\boldsymbol{a}=[7/22, 3/22, 6/11]^T$ also provide an example of three admissible solutions to the system obtained by fixing the coefficients $c_1,\ldots,c_{11}$. Finally, an example with four admissible solutions is obtained by considering $\boldsymbol{\theta}_A$ with initial condition $\boldsymbol{a}=[1/4, 3/20, 3/5]^T$, shown in Figure~3 of the main text. We conclude that parameter sets equivalent to $\boldsymbol{\theta}_A$ vary as the initial condition is changed.

\section{Incorporating the initial condition in the structural identifiability analysis when two states share the same velocity}\label{SI:Sec:IC_v2=v3}

In this section, we study the structural identifiability of the process measuring the total particle density incorporating the initial condition $n_s(x,0)=a_sf(x)$, $s=1,2,3$, if two states share the same velocity ($v_2=v_3=v$). The initial-condition input-output equations obtained in Sections~4.2-4.3 of the main text still hold, but the identifiable coefficients $c_9,c_{10},c_{11}$ for $v_2=v_3$ become
\begin{equation}
\begin{aligned}
\hat c_9&=a_1(v_1-v)+v
\\
\hat c_{10}&=a_1v_1^2+(1-a_1)v^2, 
\\
\hat c_{11}
&=(v_1-v)
\left(
-\lambda_1a_1
+\lambda_2a_2p_{21}
+\lambda_3(1-a_1-a_2)p_{31}
\right).
\end{aligned}
\end{equation}
The coefficient $c_9$ identifies $a_1$, while $\hat c_{10}$ does not add any information. Moreover, since $v_1,v,\lambda_1$ are identifiable, simplifying $\hat c_{11}$, we obtain the identifiable parameter combination 
\begin{equation}\label{Eq:k4}
    k_4=(1-a_1)\lambda_3 p_{31} + a_2(\lambda_2 p_{21} - \lambda_3 p_{31}).
\end{equation}
We conclude that the model is not structurally identifiable as the four identifiable parameter combinations $k_1,k_2,k_3,k_4$ are not sufficient to identify all six model parameters $\lambda_2,\lambda_3,p_{12},p_{21},p_{31},a_2$.

\section*{Declarations}

For the purpose of open access, the author has applied a CC BY public copyright licence to any author accepted manuscript arising from this submission.

\singlespacing

%\bibliographystyle{plainnat} 
% Bibliography name
%\bibliography{}
%\linespread{0.95}
\selectfont\bibliography{sn-bibliography_supp}